\documentclass[a4paper,12pt]{article}
\usepackage[margin=2.5cm]{geometry}
\usepackage{setspace}
\usepackage[T1]{fontenc}
\usepackage[latin9]{inputenc}
\usepackage{fancyhdr}
\usepackage[titletoc,title]{appendix}
\usepackage{graphicx}
\usepackage{soul,xcolor}
\setulcolor{red}
\setstcolor{red}
\usepackage[super,nomove]{cite}
\usepackage[font=small,labelsep=period,font=onehalfspacing]{caption}
\usepackage{subcaption}
\usepackage{epstopdf}
\usepackage{floatrow}
\usepackage{pdflscape}
\usepackage{tikz}
\usepackage{morefloats}
\usepackage[section]{placeins}
\usepackage{tikz}
\usepackage{amsmath}
\usepackage{amssymb}
\usepackage{multirow}
\usepackage[nottoc]{tocbibind}
\usepackage{color}
\usepackage{authblk}
\usepackage{framed}
\usepackage{lineno}
\usepackage{nomencl}
\usepackage{array}
\usepackage{tocloft}
\usepackage{afterpage}
\usepackage{floatrow}
\usepackage{sectsty}
\usepackage{indentfirst}

\RequirePackage{ifthen}
\renewcommand{\nomgroup}[1]{%
\ifthenelse{\equal{#1}{A}}{\item[\textit{Roman letters}]}{%
\ifthenelse{\equal{#1}{B}}{\item[\textit{Greek letters}]}{
\ifthenelse{\equal{#1}{C}}{\item[\textit{Subscripts}]}{}}}}

\usepackage{multicol} 
\renewcommand*\nompreamble{\begin{multicols}{2}}
\renewcommand*\nompostamble{\end{multicols}}

\makeatletter
\renewcommand\@biblabel[1]{\textsuperscript{#1}}

\makeatother
\addtocontents{lof}{~\hfill\textbf{Page}\par}

\makeatletter
\renewenvironment{titlepage}
    {%
      \if@twocolumn
        \@restonecoltrue\onecolumn
      \else
        \@restonecolfalse\newpage
      \fi
      \thispagestyle{plain}
      \setcounter{page}\@ne
    }%
    {\if@restonecol\twocolumn \else \newpage \fi
     \if@twoside\else
        \setcounter{page}\@ne
     \fi
    }
\makeatother

\begin{document}

\begin{titlepage}

\center

\vspace*{1.2cm}
{\large \bfseries Reduced energy consumption in stirred vessels \\ by means of fractal impellers}\\[1.2cm]

{\normalsize S. Ba\c{s}bu\u{g}, G. Papadakis and J. C. Vassilicos }\\
{\small\itshape Department of Aeronautics, Imperial College London, SW7 2AZ London, UK}
\\[1.2cm]


\onehalfspacing
\begin{abstract}

\noindent Earlier studies\cite{Steiros,Basbug} have shown that the
power consumption of an unbaffled stirred vessel decreases
significantly when the regular blades are replaced by fractal ones. In
this paper, the physical explanation for this reduction is
investigated using Direct Numerical Simulations at $Re=1600$.
The gaps around the fractal blade perimeter create jets that penetrate inside the recirculation zone in the wake and break up the trailing vortices into smaller ones.
This affects the time-average recirculation pattern on the suction side. The volume of the separation region is $7\%$ smaller in the wake of the fractal
blades. The lower torque of the fractal impeller is equivalent to a
decreased transport of angular momentum; this difference stems from
the reduced turbulent transport induced by the smaller trailing
vortices. The major difference in the turbulent dissipation is
seen in the vicinity of trailing vortices, due to fluctuations of
velocity gradients at relatively low frequencies.

\vspace{0.2cm}


\end{abstract}

\end{titlepage}

\setcounter{page}{2}

\doublespacing
\section*{Introduction}

Stirred vessels are employed in a wide range of mixing applications in
the chemical, pharmaceutical and process
industries.\cite{Paul2004a} Reducing power consumption and/or
increasing mixing quality improves process efficiency. Modification of
blade design has been considered as a means to achieve this
objective. Indeed this has been the subject of extensive research in
the past decades,
\cite{Jaworski2001,Kumaresan2006,Wu2006,Georgiev2008,Vasconcelos2000,Trivellato2011}
but these efforts yielded only modest
improvements.\cite{Vasconcelos2000,Trivellato2011}

Steiros et al.\cite{Steiros} continued work in this direction and
proposed a promising new impeller design. They introduced blades of
fractal shape and compared their performance to that of regular
blades. The application of fractal geometry to blade design was
inspired by results obtained in other areas over the past ten years,
summarized by the aforementioned authors\cite{Steiros} and in our
previous paper.\cite{Basbug} More specifically, Steiros et
al.\cite{Steiros} performed shaft torque measurements to determine the
power consumption of four-bladed radial impellers in an unbaffled tank
at $Re=1-2\times10^5$, where $Re=ND^2/\nu$ is based on the rotational
speed $(N)$, impeller diameter $(D)$ and kinematic viscosity of the
fluid $(\nu)$. Their results demonstrated that the fractal-1 impeller,
that uses blades with one fractal iteration (seen in Figure
\ref{fig:cp_fra}) has $11-12\%$ reduced power consumption compared to
the impeller with regular blades (seen in Figure
\ref{fig:cp_reg}). When the fractal-2 impeller was employed (with two
fractal iterations, see Figure 2 in Steiros et al.\cite{Steiros}),
this difference increased to $17-20\%$. The authors also measured the
pressure distribution on both sides of the blades using pressure
transducers. They found that the center of pressure for a fractal
blade is located radially further away from the shaft compared to a
regular one. Consequently it was shown that the decreased torque on
the fractal blade is due to the lower net pressure force applied onto
the blade and \textit{not} due to the reduced moment arm length. Since
both types of blades have equal frontal area by construction, it was
concluded that the fractal blade has a lower drag
coefficient. Furthermore, they tested a two-bladed impeller, where one
regular and one fractal blade were mounted 90\textdegree\ apart. They
conducted experiments for both rotational directions, whereby the
regular blade was immersed in the wake of the fractal blade and vice
versa. They measured exactly the same torque in both experiments. This
finding suggests that the wake interaction was \textit{not} the reason
for the reduced drag coefficient of a fractal blade. However, their
results did not identify and explain clearly the mechanism which leads
to the reduced torque.

In the present study, we use Direct Numerical Simulations (DNS) to
conduct a detailed analysis of the flows around both impeller
types. We aim to answer the following open questions: Why does the
fractal impeller require reduced torque and power consumption with
respect to a regular one with equal frontal area? How is this related
to the flow patterns in the wake? The physical understanding from
such an investigation can lead to further improvements in impeller
design in the future.

We have performed DNS of the flow field inside an unbaffled stirred
vessel with four-bladed radial impellers at $Re=320$ and
$Re=1600$. The DNS approach produced high fidelity, well-resolved data
of velocity and pressure in space and time. Such time-accurate and
comprehensive database allows a detailed analysis of the wake flow
patterns, and their effect on the pressure field around the impeller
and the distribution of energy dissipation in the whole tank. There
are few other DNS studies of the flow in baffled\cite{Gillissen2012}
and unbaffled\cite{Verzicco2004} stirred tanks, but with different
goals to those of the present work.

The remainder of this paper is organized as follows: In the following
section, we introduce the flow configuration and describe briefly the numerical methodology employed. This is followed by a comparison
between experimental and numerical results for validation, and a
comparative summary of the power consumption. The pressure
distribution on regular and fractal blades and its relation to the
flow is presented and analysed. Differences in transport of angular
momentum and in energy dissipation between the two impeller types are
investigated. We close the paper with a summary and main conclusions.

\section*{Flow configurations examined}

\begin{figure}[ht]
  \centering   
    \begin{subfigure}[t]{0.45\textwidth}
    \centering   
        \includegraphics[width=0.7\textwidth]{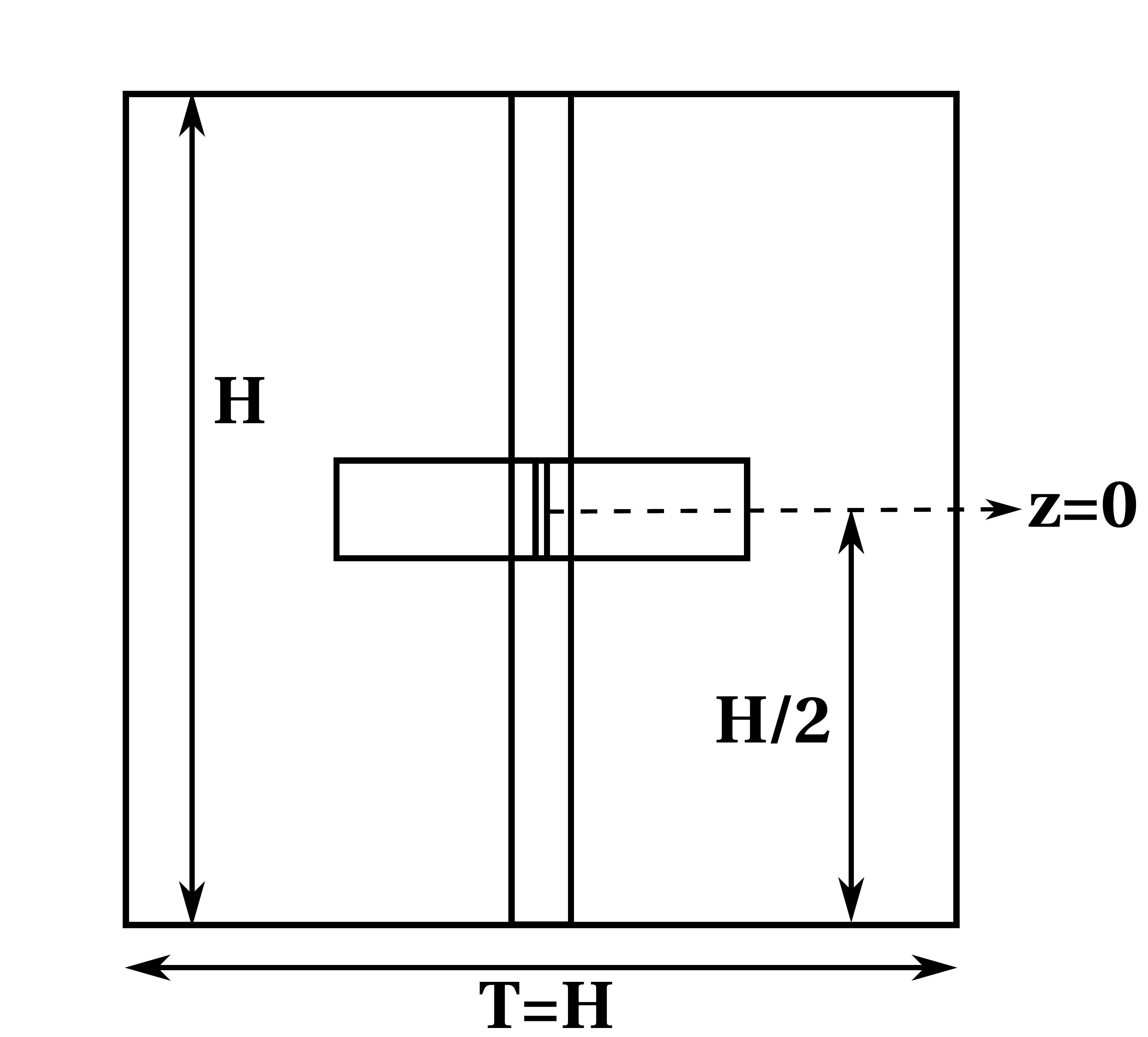}
        \caption{}
        \label{fig:tank_a}
    \end{subfigure}  
    \begin{subfigure}[t]{0.45\textwidth}
    \centering 
        \includegraphics[width=0.78\textwidth]{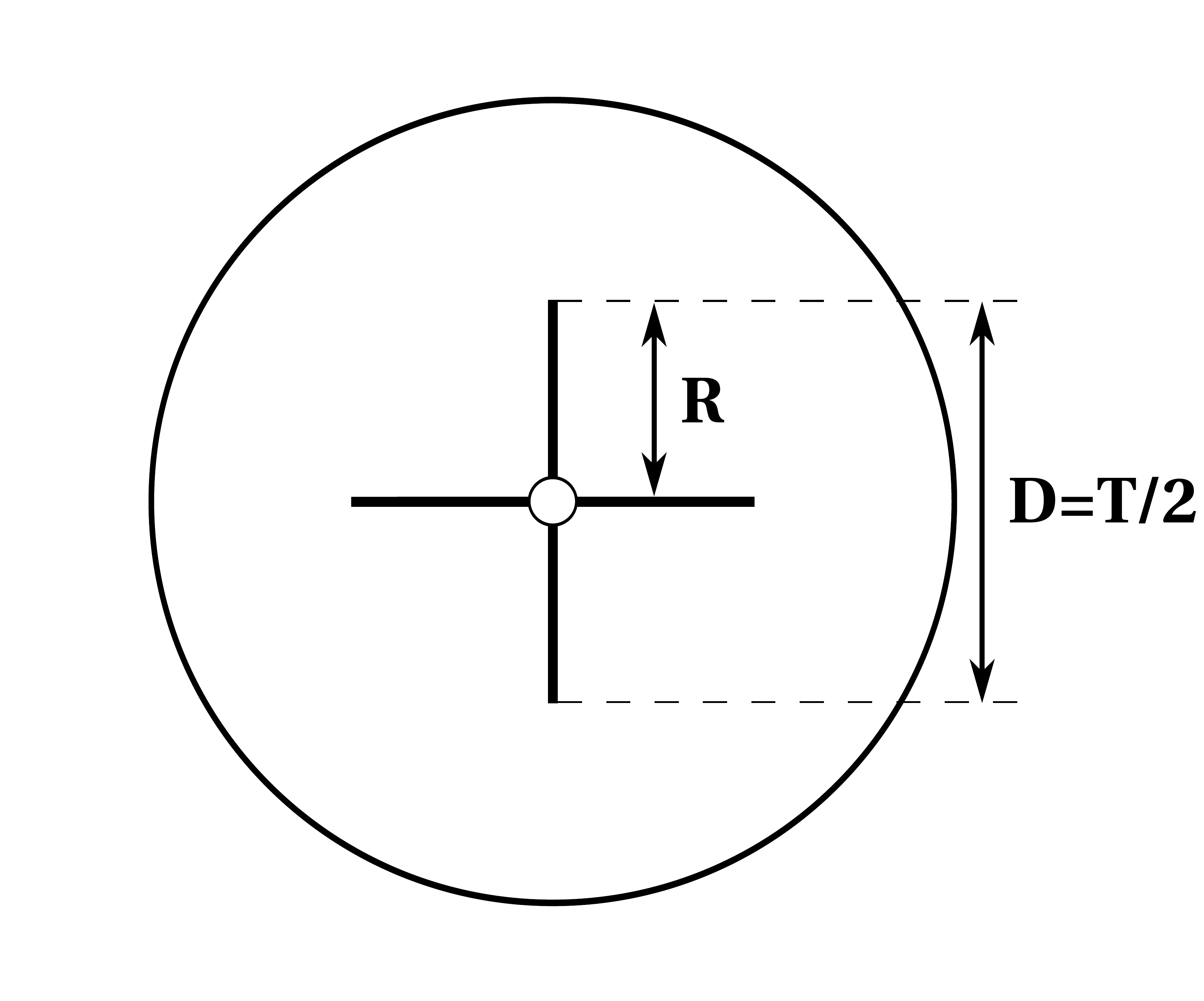}
        \caption{}
        \label{fig:tank_b}
    \end{subfigure}  
  \caption{Tank geometry and dimensions. (a) Vertical view along the
    axis, (b) horizontal (top) view at the mid-height.}
  \label{fig:tank}
\end{figure}

The stirred tank used in the present study has a cylindrical geometry
of equal height and diameter, with a four-bladed impeller located at
the mid-height. Figures \ref{fig:tank_a} and \ref{fig:tank_b} show
vertical and horizontal views of the vessel and the impeller. The
blade height and blade thickness are $1/10$ and $1/100$ of the tank
diameter, respectively. Two types of impeller blades with equal
frontal area were employed in the present work, referred to below as
regular and fractal blades, illustrated in Figures \ref{fig:cp_reg}
and \ref{fig:cp_fra} respectively. These blades are the same ones used
in our previous DNS study.\cite{Basbug} They also match some of the
blades in the experimental study of Steiros et al.\cite{Steiros}

As seen in Figure \ref{fig:cp}, the radius of the fractal impeller is
not uniform. The furthest tip is located at $r/R=1.1$, where $R$ is
the constant radius of the regular impeller. For normalization
purposes, the dimensions of the regular impeller are used throughout
the paper.

\section*{Numerical methodology} 

The incompressible in-house code ``Pantarhei'' was used to obtain the
DNS results presented in the paper. The code is based on the
finite-volume method and was previously employed for DNS of the flow
around an airfoil,\cite{Thomareis2017} the flow past a single square
grid-element\cite{Paul2017} and the flow in an unbaffled stirred
vessel.\cite{Basbug} More details can be found in the latter reference
which deals with the same configuration as the present study. The
Navier-Stokes equations were solved in a rotating reference frame,
fixed to the impeller. In this frame, the momentum equations take the
form:
\begin{equation}
\frac {\partial\rho\vec{v}}{\partial t}+\nabla \cdot{(\rho\vec{v}\otimes\vec{v})}=
-\nabla p+\nabla\cdot\boldsymbol{\tau}-\rho[\: \vec{\omega} \times (\vec{\omega} \times \vec{r})+2\vec{\omega} \times \vec{v} \:],
\label{eq:frame}
\end{equation}
where $\vec{v}$ is the instantaneous velocity vector in the rotating
frame, $\boldsymbol{\tau}$ is the viscous stress tensor and
$\vec{\omega}$ denotes the angular velocity of the frame. The origin
of the coordinate system is located on the impeller axis at the
mid-height of the vessel. The position vector with respect to the
origin is denoted by $\vec{r}$. Vector $\vec{\omega}$ points in the
axial direction and has magnitude $\Omega=2 \pi N$. The last two terms
of Equation \ref{eq:frame} represent the centrifugal and Coriolis
forces respectively, and were treated as source terms. A prescribed
velocity (equal to $\omega r_w$, where $r_w$ is the radial distance of
a wall point to the axis of the vessel) was imposed on all external
walls to represent the relative motion with respect to the
impeller. The top boundary was also treated as a solid wall, hence
free-surface depression was not considered. In Basbug et
al.\cite{Basbug} the results of the code in terms of mean and rms
velocities were validated against those of Verzicco et
al.\cite{Verzicco2004} and Dong et al.\cite{Dong1994}

The flow inside the vessel was simulated at two Reynolds numbers,
$Re=320$ and $Re=1600$. For $Re=320$, the grids consisted of
$13\times10^6$ and $21\times10^6$ cells for the regular and fractal
impellers, respectively. For $Re=1600$, we used $60\times10^6$ and
$70\times10^6$ cells, respectively. Details about the grid resolution
and convergence study can be found in Basbug et al.\cite{Basbug}

\section*{Comparison with experimental results}

The numerical results at $Re=1600$ are compared with data acquired by 
means of phase-locked planar particle image velocimetry (PIV) at $Re=1.5 \times
10^5$. The experiments were performed using blades of the same shape
in an unbaffled tank of the same dimensions. The only difference was
that the tank was hexagonal instead of cylindrical, but this
difference is not expected to affect the results in the near impeller
region. More details on the stirred tank setup and experimental
technique can be found in Steiros et al.\cite{Steiros2017}

\begin{figure}[ht] 
  \centering   
\includegraphics[width=0.6\textwidth]{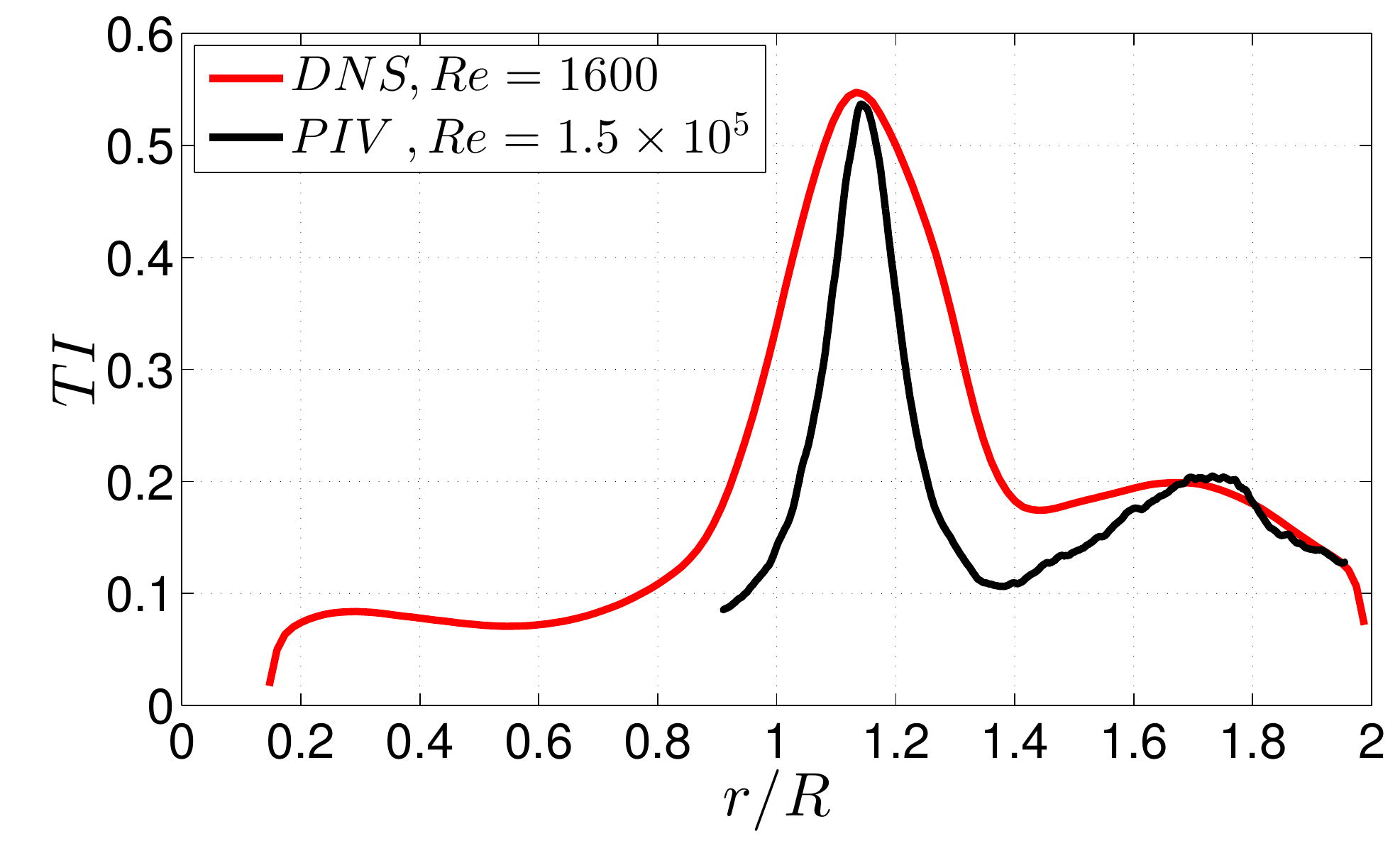}
  \caption{Profiles of turbulence intensity ($TI$) along a radial line
    30\textdegree\ behind the regular impeller, at the axial position
    where the mean azimuthal vorticity is highest in the upper
    trailing vortex core.}
\label{fig:ti_piv}
\end{figure}

The instantaneous velocity vector in the absolute reference frame is
denoted by $\vec{u}$ $(\vec{u}=\vec{v}+\vec{\omega} \times \vec{r})$
and can be decomposed in mean and fluctuating components as
$\vec{u}=\vec{U}+\vec{u'}$; the same decomposition is applied to
$\vec{v}$ as well later in the text. The azimuthal velocity component
was not accessible in the aforementioned experiment. Using the radial
and axial velocity fluctuations only, $u_r'$ and $u_z'$ respectively,
the turbulence intensity ($TI$) is defined as follows:
\begin{equation}
TI= \frac{\sqrt{ \left< u_r'^{\:2} \right> + \left< u_z'^{\:2} \right>
}}{U_{tip} } \:\:,
\label{eq:ti_piv}
\end{equation}
where angular brackets $\left < \right >$ represent the time-averaging
operation and $U_{tip}=\Omega R$ is the blade tip velocity. Profiles
of $TI$ obtained numerically and experimentally are compared in Figure
\ref{fig:ti_piv} along a radial line 30\textdegree\ behind the regular
impeller, at an axial position where the center of the upper trailing
vortex core is located, where the mean azimuthal vorticity, i.e. $
(\nabla \times \vec{U}) \cdot\hat{e}_{\theta}$, is at its
highest. This point is preferred to a fixed axial coordinate since the
location of vortex cores depends on $Re$.

As seen in Figure \ref{fig:ti_piv}, the two curves are in qualitative
agreement despite the large difference in $Re$ (two orders of
magnitude). The peaks of $TI$ indicate the locations of the vortex
cores and their radial positions (between $r/R=1.1-1.2$) are in
agreement. A second, but less prominent peak, appears around $r/R=1.7$
due to the wake of the preceding blade, and in this region the curves
collapse. Yoon et al.\cite{Yoon2005} investigated the Reynolds number
scaling of the flow in an unbaffled tank stirred with a Rushton
turbine using phase-locked stereoscopic PIV measurements for a range
of $Re$ values between $4\times10^3$ and $78\times10^3$. They reported
that the vortex core diameter decreases as $Re$ rises. The results
shown in Figure \ref{fig:ti_piv} are in accordance with their
observation, the peaks of $TI$ become sharper and narrower with
increasing $Re$.

Raju et al.\cite{Raju2005} conducted a similar study (using the same 
configuration as Yoon et al.\cite{Yoon2005}) and measured the
rms of the components of $\vec{u'}$ in a plane normal to the radial
direction close to a blade tip. They reported that the rms values 
(averaged over the measurement plane) decrease slowly with increasing $Re$, 
from $Re=4\times10^3$ to ca. $Re=4\times10^4$. For larger $Re$, 
the rms values remained constant. The experimental results presented in Figure
\ref{fig:ti_piv} were obtained for a $Re$ higher than the aforementioned threshold
for $Re$-independence. The moderately higher level of $TI$ at
$Re=1600$ is also in accordance with the findings of Raju et
al.\cite{Raju2005} In conclusion, the present numerical results are in
qualitative agreement with experimental findings. The observed
discrepancies are attributed to the large difference in Reynolds
number.

\section*{Power consumption for the two blade types}

The power, $P$, drawn by the impeller can be computed using two
different approaches. In the first approach, the torque of the
pressure and skin friction forces applied on the impeller and shaft
surfaces is integrated to compute the total impeller torque,
$T_{imp}$, which is then multiplied with the angular velocity, so
$P=T_{imp} \: \Omega$. In the second approach, the dissipation of the
total kinetic energy, $\varepsilon_K$, is integrated over the tank
volume. The dissipation is defined as $\varepsilon_K=2\nu
s_{ij}s_{ij}$, where $s_{ij}$ is the strain-rate tensor,
$s_{ij}=\frac{1}{2} \left( \frac{\partial u_i}{\partial
  x_j}+\frac{\partial u_j}{\partial x_i} \right)$. This definition
includes the contributions of both the turbulent dissipation and the
dissipation due to the mean velocity gradients. Since the tank is a
closed volume, the source and sink of energy must balance on
average. Therefore, the power injected by the impeller must equal the
total dissipation, when both quantities are averaged over a long
period of time.

The power number is defined as $N_p=P/(\rho N^3 D^5)$. Table
\ref{table:all_power} provides the values of $N_p$ based on the power
input of the impeller and the volume integral of dissipation, averaged
over 80 revolutions for $Re=320$ and for 120 revolutions for
$Re=1600$ (also the other mean quantities presented in this paper
were averaged for the same duration as stated for $N_p$ at the 
corresponding $Re$). The observed imbalance between the two approaches 
varies between $2.2\%$ and $4.2\%$. It was also shown in the previous
study\cite{Basbug} that the torque applied by the impeller is in very
good agreement with the integral torque applied by the tank walls,
with discrepancies between $0.15\%$ and $1.5\%$. This suggests that
the approach based on the impeller torque is a more reliable method
for the calculation of the power consumption compared to the volume
integration of dissipation.

\begin{table}[!ht] 
\caption{Averaged power numbers ($N_p$) for $Re=320$ and $1600$, computed using the impeller power $(P)$ and integral of dissipation. Reg. stands for regular and Fr. for fractal impeller.}
\begin{tabular}{ l || c | c | c | c} 
 Blade type, Re        & Reg., $320$  &  Fr., $320$ & Reg., $1600$ & Fr., $1600$  \\ \hline 
 Impeller power        & $ 2.32  $   &  $ 2.32  $  &  $ 1.59  $  & $ 1.47  $ \\ 
 Integral dissipation  & $ 2.27  $   &  $ 2.25  $  &  $ 1.53  $  & $ 1.41  $ \\  
 Imbalance             & $ 2.2\% $   &  $ 3.1\% $  &  $ 3.8\% $  & $ 4.2\% $ \\
\end{tabular}
\label{table:all_power}
\end{table}

The results on the impeller power show that the power consumption is
equal for both types of impellers at $Re=320$, with $N_p=2.32$ (see
Table \ref{table:all_power}). In fact, the contribution of the shear
stress to the impeller torque is only $4\%$ for the regular impeller,
whereas it is $8\%$ for the fractal impeller at this $Re$. This
difference is expected because a fractal blade has ca. twice the
perimeter length of a regular blade, as seen in Figure
\ref{fig:cp}. If only pressure forces were taken into account, we
would obtain a lower $N_p$ with the fractal impeller by ca. $4\%$,
which is compensated by the increased skin friction drag.

At $Re=1600$, the contribution of the shear stress is an order of
magnitude lower than at $Re=320$, and it therefore has no significant
influence on the power consumption. Also at $Re=1600$, the numerical
results suggest a power reduction of ca. $8\%$ when the regular blades
are replaced with fractal blades. This difference is in accordance
with the aforementioned experimental results\cite{Steiros} obtained at
much higher $Re$. In the following sections, we endeavour to provide a
physical explanation for this behaviour, mainly focusing on the
results at $Re=1600$. In order to arrive at a coherent picture, we
consider the pressure distribution over the blades, the transport of
angular momentum and how it is affected by vortical wake structures,
the distribution of dissipation in the wake and the whole vessel, as
well as the frequency content of dissipation.

\section*{Pressure distribution on blade surfaces}

Considering that both regular and fractal impellers rotate with the
same angular velocity, any difference in the power is a consequence of
a different torque on the impeller. Therefore, we start the analysis
with an investigation of pressure distribution on the blade
surfaces. The time-averaged pressure difference between the suction
and pressure sides is normalized to define a local pressure
coefficient, $C_p$, as follows:

\begin{equation}
C_p=\frac{\Delta p}{0.5 \, \rho \, U_{tip}^2 } \:.    
\end{equation}  

\begin{figure}[ht]
  \centering   
    \begin{subfigure}[c]{0.53\textwidth}
    \centering   
        \includegraphics[width=\textwidth]{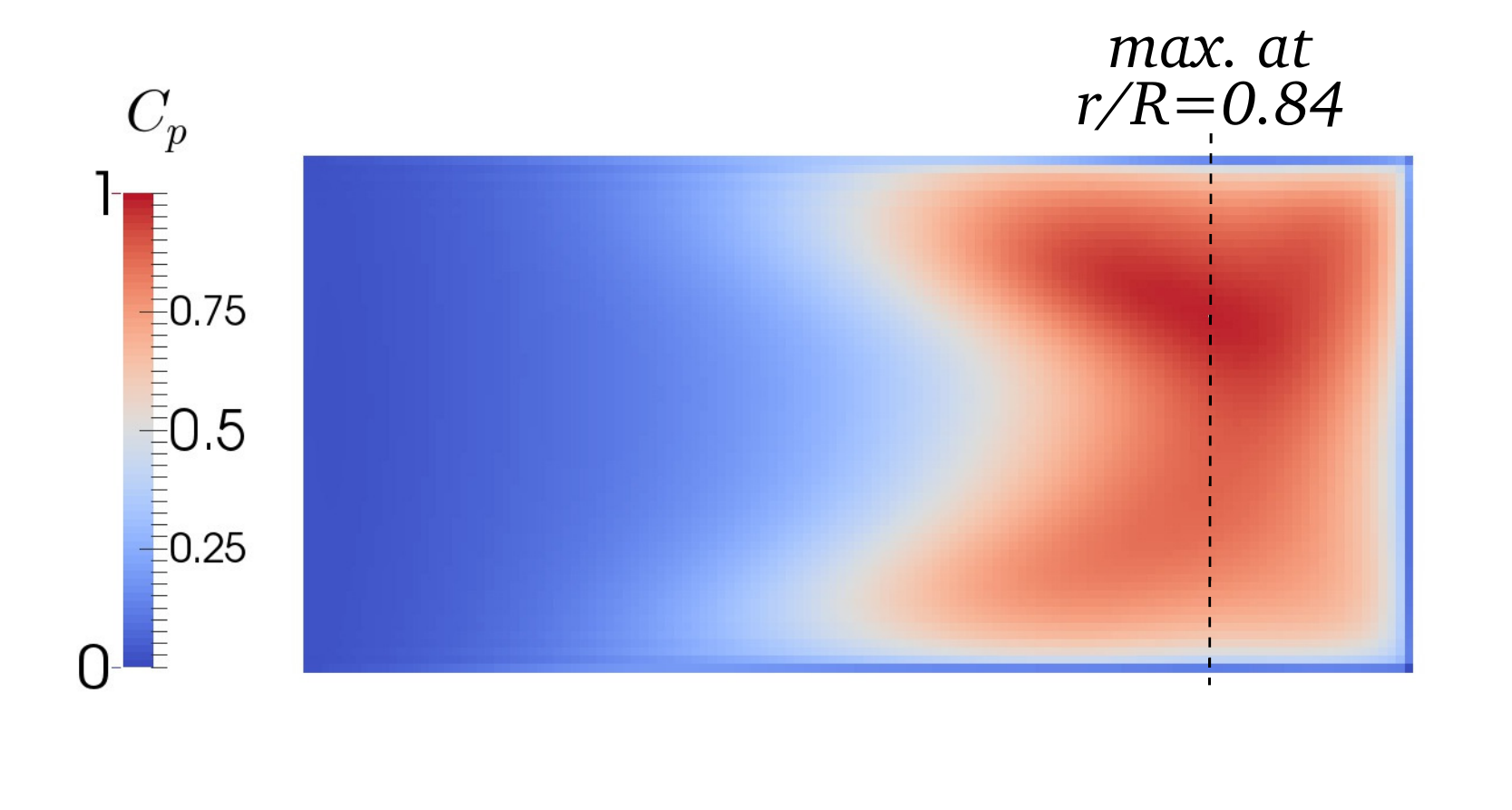}
        \caption{}
    \label{fig:cp_reg}    
    \end{subfigure}  
    \begin{subfigure}[c]{0.46\textwidth}
    \centering 
        \includegraphics[width=\textwidth]{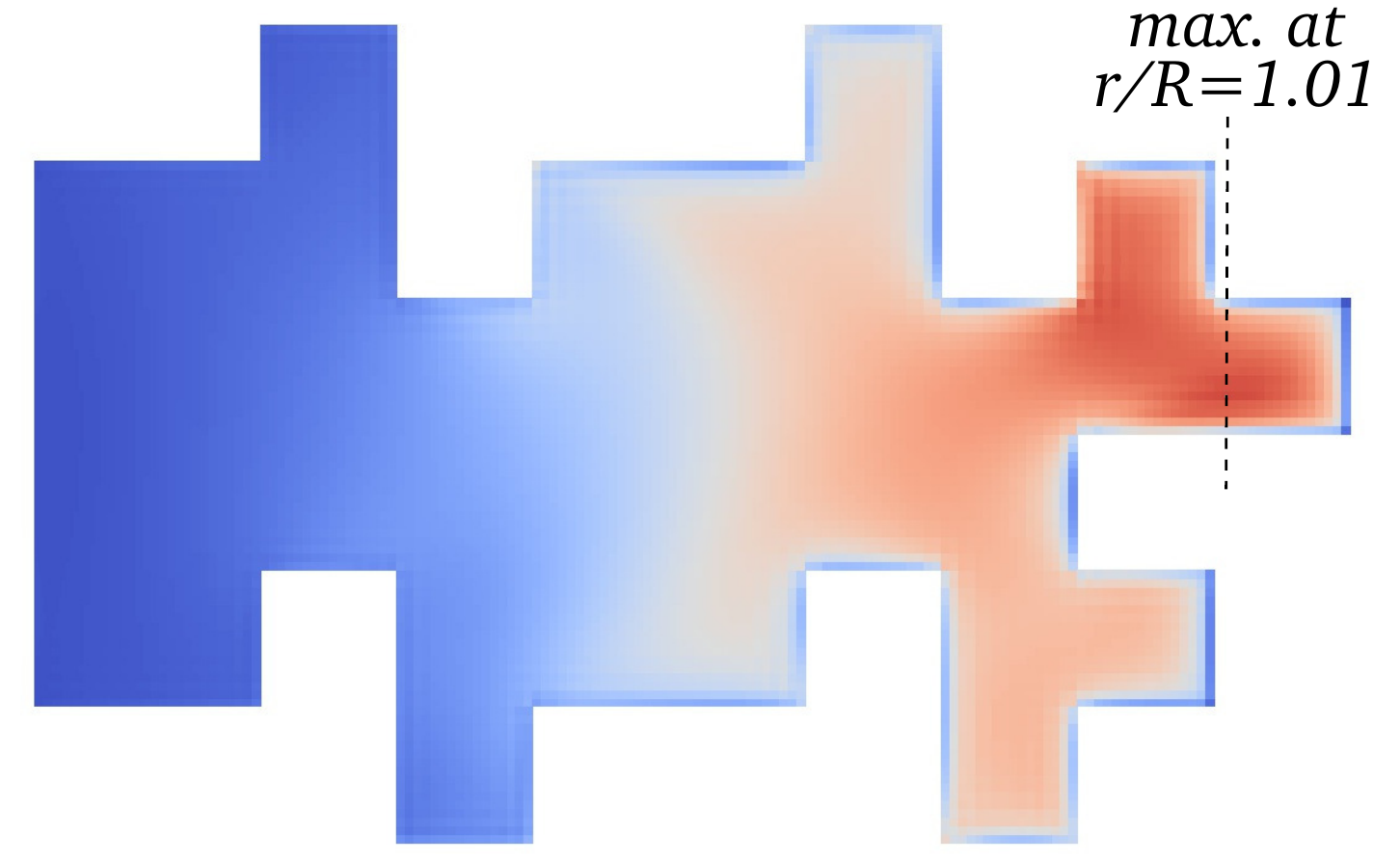}
        \caption{}
    \label{fig:cp_fra}    
    \end{subfigure}  
  \caption{$C_p$ distribution on the surface of (a) regular and (b)
    fractal blade. The radial positions of maximum $C_p$ are indicated
    with vertical dashed lines.}
\label{fig:cp}
\end{figure}

The distribution of $C_p$ over the blades surface is illustrated in
Figure \ref{fig:cp}, and qualitatively it agrees well with the
experimental results.\cite{Steiros} The maximum value for the regular
blade is close to $1$, whereas for the fractal blade it reaches up to
$0.9$. In the experiments\cite{Steiros} with $Re=1-2\times10^5$, the
maximum value was less than $0.75$. It is expected that this quantity
will be lower at higher $Re$, because $N_p$ is lower as well. The
radial positions of $C_p$ maxima are denoted with vertical dashed
lines in Figure \ref{fig:cp}. These are located at $r/R=0.84$ 
and $1.01$ for the regular and fractal blades, respectively. From the $C_p$ distributions reported in the aforementioned experimental study, it appears that the maxima are located in similar positions, but the exact locations were not stated.

From the surface integral of $C_p$, the drag coefficient can be
computed as $C_D=( \int{C_p \: dA} )/ \int{dA}$. The values of $C_D$
from the present DNS study and the experiments\cite{Steiros} at
$Re=10^5$ are listed in Table \ref{table:all_drag}. Note that $C_D$ is
based on pressure forces only, i.e. viscous forces are excluded. The
center of pressure, $CoP$, is also included in Table
\ref{table:all_drag}. It is defined as the net torque divided by the
net force applied on the blade, and it is normalized with the impeller
radius, $R$. As expected, $C_D$ decreases with the increasing $Re$, a
behaviour similar to that of $N_p$. Moreover, $CoP$ moves radially
further from the axis as $Re$ rises. At every $Re$, the fractal blade
has a lower $C_D$ than the regular blade. At $Re=320$, the larger
value of $CoP$ and the more significant viscous forces on the fractal
blade compensate for the lower $C_D$, so that both blades apply the
same torque on the fluid. At the higher $Re$ numbers, the reduced
pressure drag force on the fractal impeller is the dominant factor
that explains the lower torque with respect to the regular impeller.

\begin{table}[ht] 
\caption{Drag coefficients of both types of blades at $Re=320$, $1600$
  and $10^5$. Reg. stands for regular and Fr. for fractal
  impeller. CoP denotes the radial position of the center of
  pressure. The values at $Re=10^5$ are taken from Steiros et
  al.\cite{Steiros}}
\begin{tabular}{ l || c | c | c | c | c | c} 
 Blade type & Reg. &  Fr. & Reg. &  Fr. & Reg. & Fr. \\  
 Re & $320$ &  $320$ & $1600$ &  $1600$ & $10^5$ (exp.) & $10^5$ (exp.) \\ \hline 
 $C_D$          & $ 0.597 $  &  $ 0.538 $  & $ 0.406 $  &  $ 0.363 $  
 &  $ 0.172 $  & $ 0.153 $ \\ 
 $CoP/R$        & $ 0.698 $  &  $ 0.728 $  & $ 0.735 $  &  $ 0.736 $  
 &  $ 0.820 $  & $ 0.844 $ \\  
\end{tabular}
\label{table:all_drag}
\end{table}

\begin{figure}[ht] 
  \centering   
    \begin{subfigure}[b]{0.5\textwidth}
    \centering 
        \includegraphics[width=\textwidth]{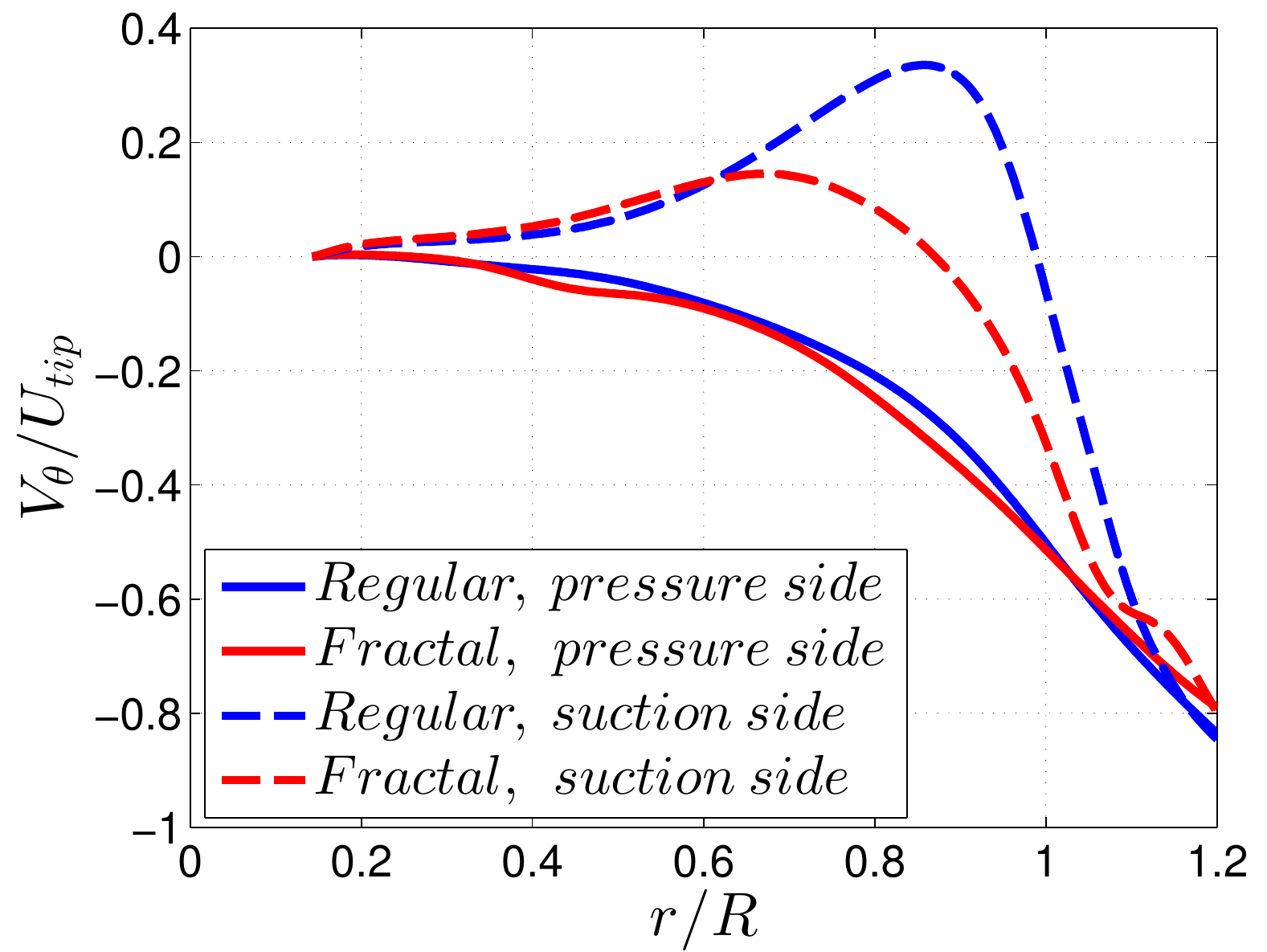}
        \caption{}
    \end{subfigure}  
        \begin{subfigure}[b]{0.40\textwidth}
    \centering   
        \includegraphics[width=0.85\textwidth]{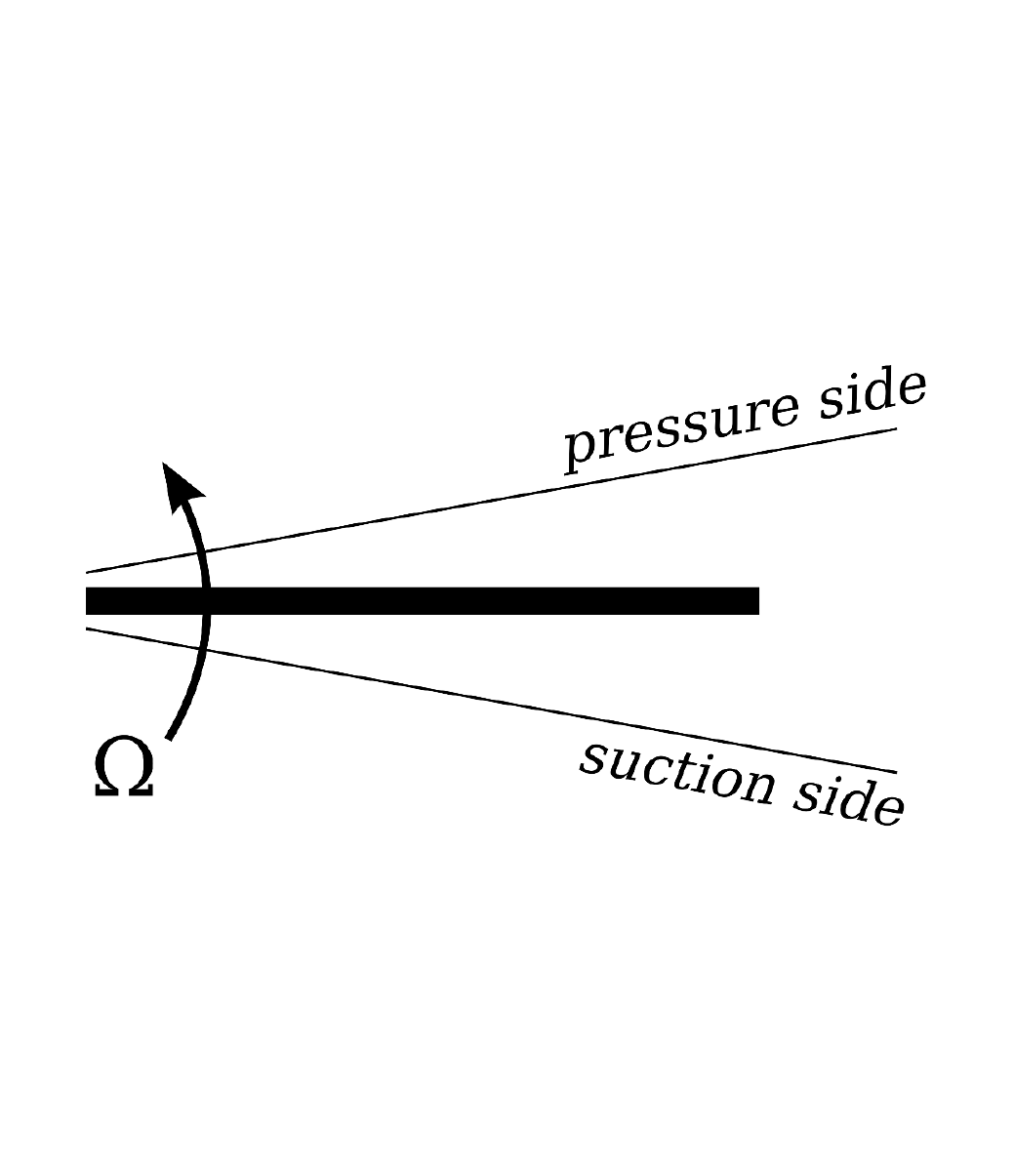}
        \caption{}
    \end{subfigure}  
  \caption{(a) Time-averaged azimuthal velocity profiles along a radial
    line 10\textdegree\ upstream of the pressure side and
    10\textdegree\ downstream of the suction side of regular and
    fractal blades, at the tank mid-height in relative frame at
    $Re=1600$. (b) Illustration of the radial lines along which the profiles were  acquired (shown here for a regular blade) and the direction of rotation.}
 \label{fig:profiles}
\end{figure}

In order to gain more insight on the observed pressure distributions,
we turn our attention to the velocity fields around the blades. Figure
\ref{fig:profiles}a shows the mean azimuthal velocity profiles
acquired along two radial lines, at 10\textdegree\ upstream of the
pressure side and 10\textdegree\ downstream of the suction side (as
depicted in Figure \ref{fig:profiles}b). These profiles were obtained
at the mid-height of the tank, in the relative reference frame, at
$Re=1600$. Since the impeller rotation is in the positive azimuthal
direction, as indicated with an arrow in Figure \ref{fig:profiles}b,
the relative azimuthal velocities ($V_{\theta}$) will have negative
values on the pressure side along the entire profile. For the radial
positions up to half of the blade radius ($r/R<0.5$), the velocity
magnitudes are less than $10\%$ of $U_{tip}$ for both types of blades
on both sides. This explains why $C_p$ values are very low on the
inner half of the blades, as seen in Figure \ref{fig:cp}.

The profiles of $V_{\theta}$ on the pressure sides of both blades
collapse (solid lines in Figure \ref{fig:profiles}a). This is
consistent with the results of experiments\cite{Steiros} using a two
bladed impeller, where one regular and one fractal blade are mounted
90\textdegree\ apart. The measured torque was the same when the
impeller rotated in either direction, immersing the regular or fractal
blade in the wake of the other. This led to the conclusion that the
wake interaction is \textit{not} the reason of the reduced drag
coefficient of a fractal blade. Here we elaborate further and show
that this is because the upstream velocity profiles are almost
independent of the type of the preceding blade. On the other hand, the
velocities downstream of the blades are significantly different
(dashed lines in figure \ref{fig:profiles}a). This may seem
contradictory if one expects that the downstream profile of a blade
must interact with the following blade. In our observation, there is a
strong radial jet along the suction side of the blades which carries
the fluid in the wake radially away. The part of the flow field
immediately upstream of the blade is mainly advected axially from the
top and bottom of the tank towards the mid-height along the shaft,
before interacting with the pressure side. Therefore, the profiles on
the pressure side are strongly affected by the angular momentum of the
fluid that is in the bulk of the flow far from the impeller. If the
bulk flow had higher angular momentum (i.e. higher positive azimuthal
velocity in absolute frame, as it is observed at increased
$Re$\cite{Nagata1975}), then the profiles of $V_{\theta}$ on the
pressure side would have a smaller magnitude (i.e. lower velocity
relative to the blade). This would lead to a weaker stagnation on the
pressure side and a lower drag coefficient, as it is seen in Table
\ref{table:all_drag} for growing $Re$.

We define the flow separation zone in the wake of a blade as the
region where $V_{\theta}$ is positive, i.e. towards the blade suction
side (see Figure \ref{fig:profiles}b). This region extends until about
$r/R=1$ in the wake of the regular impeller (blue dashed line in
Figure \ref{fig:profiles}a). Moreover, the magnitude of $V_{\theta}$
reaches a maximum value of $0.34 \, U_{tip}$. On the other hand, in
the wake of the fractal impeller $V_{\theta}$ becomes negative again
around $r/R=0.88$ and the maximum magnitude is only $0.16\, U_{tip}$,
indicating a smaller and weaker separation zone.

\begin{figure}[ht]
  \centering   
    \begin{subfigure}[c]{0.18\textwidth}
    \centering   
        \includegraphics[width=\textwidth]{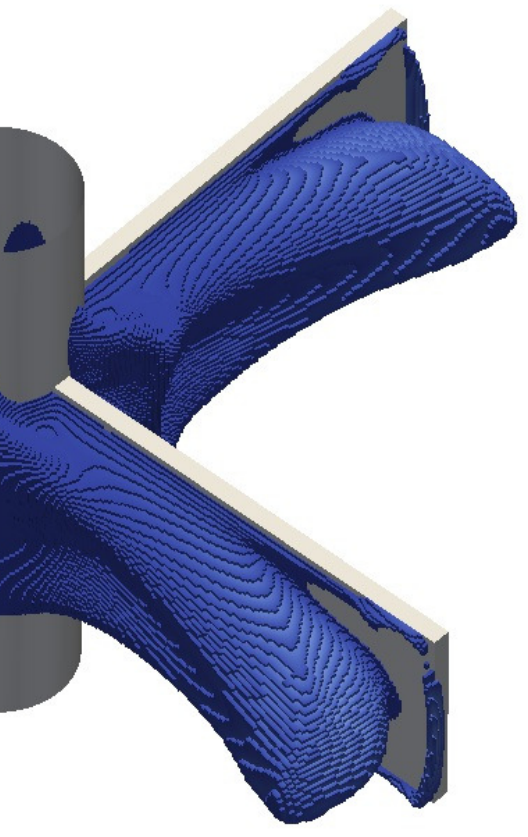}
        \caption{}
    \end{subfigure}  
    \begin{subfigure}[c]{0.3\textwidth}
    \centering 
        \includegraphics[width=\textwidth]{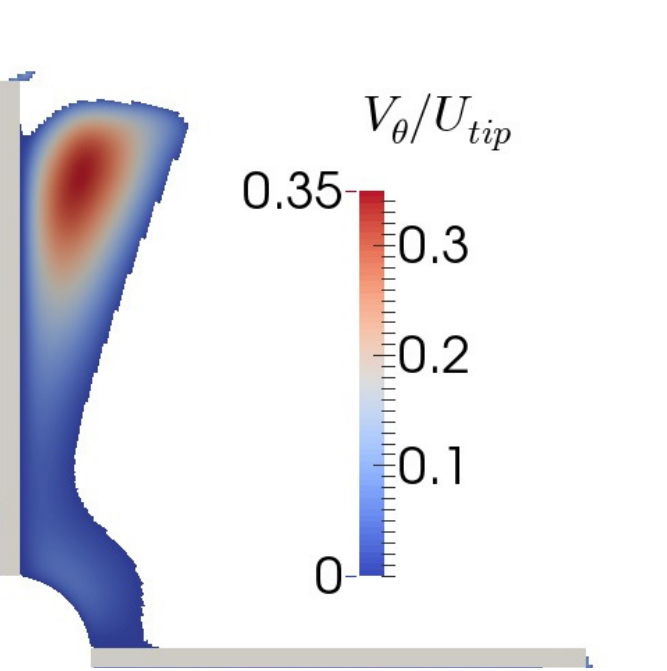}
        \caption{}
    \end{subfigure}  
    \begin{subfigure}[c]{0.18\textwidth}
    \centering   
        \includegraphics[width=\textwidth]{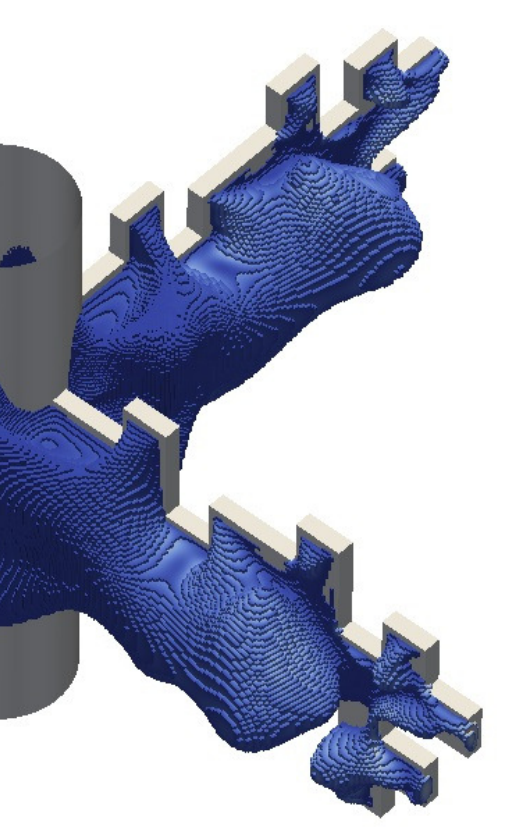}
        \caption{}
    \end{subfigure}  
    \begin{subfigure}[c]{0.3\textwidth}
    \centering 
        \includegraphics[width=\textwidth]{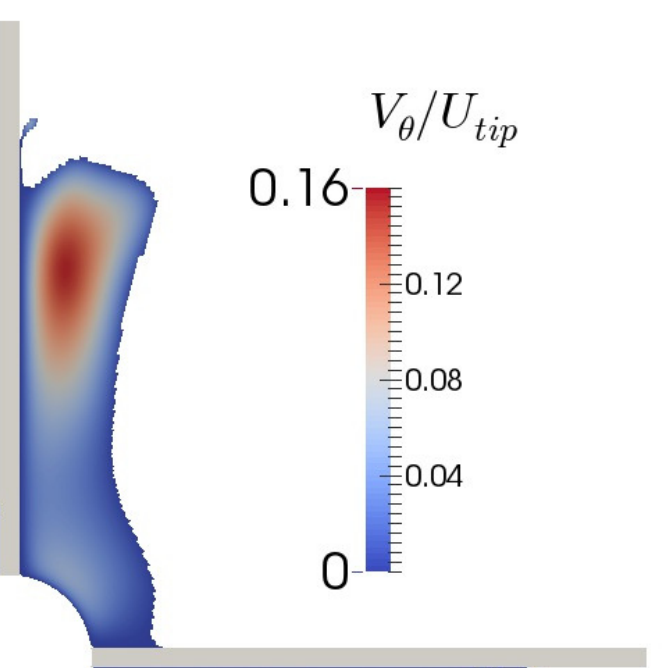}
        \caption{}
    \end{subfigure}  
  \caption{Flow separation regions (where $V_{\theta}>0$) behind the
    blades at $Re=1600$, (a) 3-D view for regular impeller , (b)
    contours of $V_{\theta}$ in a cross section of the volume shown in
    (a), (c) 3-D view for fractal impeller, (d) contours of $V_{\theta}$
    in a cross section of the volume shown in (c).}
  \label{fig:seperation}
\end{figure}

The separation zone as defined above is illustrated in Figures
\ref{fig:seperation}a and \ref{fig:seperation}c in a three dimensional
view in the wakes of regular and fractal blades, respectively. Taking
all four blades into account, the size of the separated region is
$14.1\%$ and $13.2\%$, respectively, of the impeller swept volume
($\pi R^2 H/10$). The larger separating zone in the wake of a regular
blade (by about $7\%$) causes a lower pressure on the suction side,
resulting in a higher form drag. At first sight, this finding seems to
contradict the results of Nedi\'c et al.\cite{Nedic2013} who performed
wind tunnel experiments with square and fractal plates placed normal
to an incoming flow. The shapes of the plates were similar to the
blades used in the present study. In this flow setting, they measured
larger $C_D$ values for the fractal plates compared to the orthogonal
plates. However, this flow setting is different to the one examined in
the present paper. While the front stagnation flow patterns may be
qualitatively similar, the wakes have markedly different
behaviour. The rotation of the blade creates a strong radial jet, that
is absent when the object is placed normal to the flow. The presence
of the radial jet changes entirely the wake properties, resulting in
different trends in the $C_D$ values.

Figures \ref{fig:seperation}b and d illustrate contours of
$V_{\theta}$ in a cross section of the separation zones at the tank
mid-height. The dark red colour indicates the location with the
highest $V_{\theta}$. This location is radially further away from the
axis for the regular blade compared to the fractal blade, and matches
with the radial location where the highest $C_p$ was observed in
Figure \ref{fig:cp}a, at $r/R=0.84$. On the other hand, the fractal
blade has the highest $C_p$ near the tip of the blade, as seen in
Figure \ref{fig:cp}b, which does not coincide with the location of the
strongest separation. This can be explained as follows: Since the tip
of the fractal blade extends radially further (up to $r/R=1.1$) than
the tip of the regular blade (up to $r/R=1$), the approaching relative
velocity is higher, leading to higher stagnation pressure on the
pressure side. This is the dominant factor that determines the
location of highest $C_p$ for the fractal blade.

\begin{figure}[ht] 
  \centering   
    \begin{subfigure}[b]{0.45\textwidth}
    \centering   
        \includegraphics[width=\textwidth]{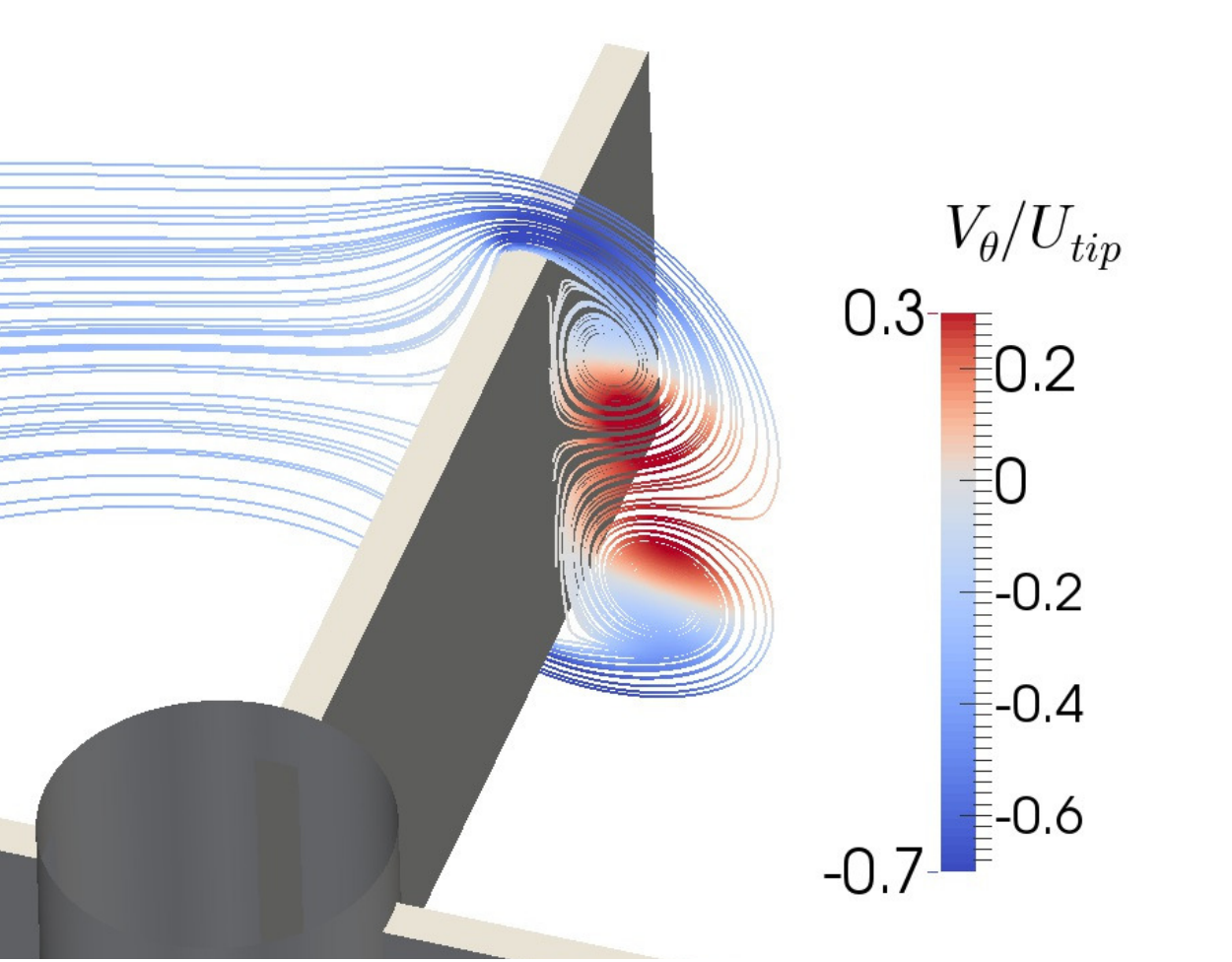}
        \caption{}
    \end{subfigure}  
    \begin{subfigure}[b]{0.45\textwidth}
    \centering 
        \includegraphics[width=\textwidth]{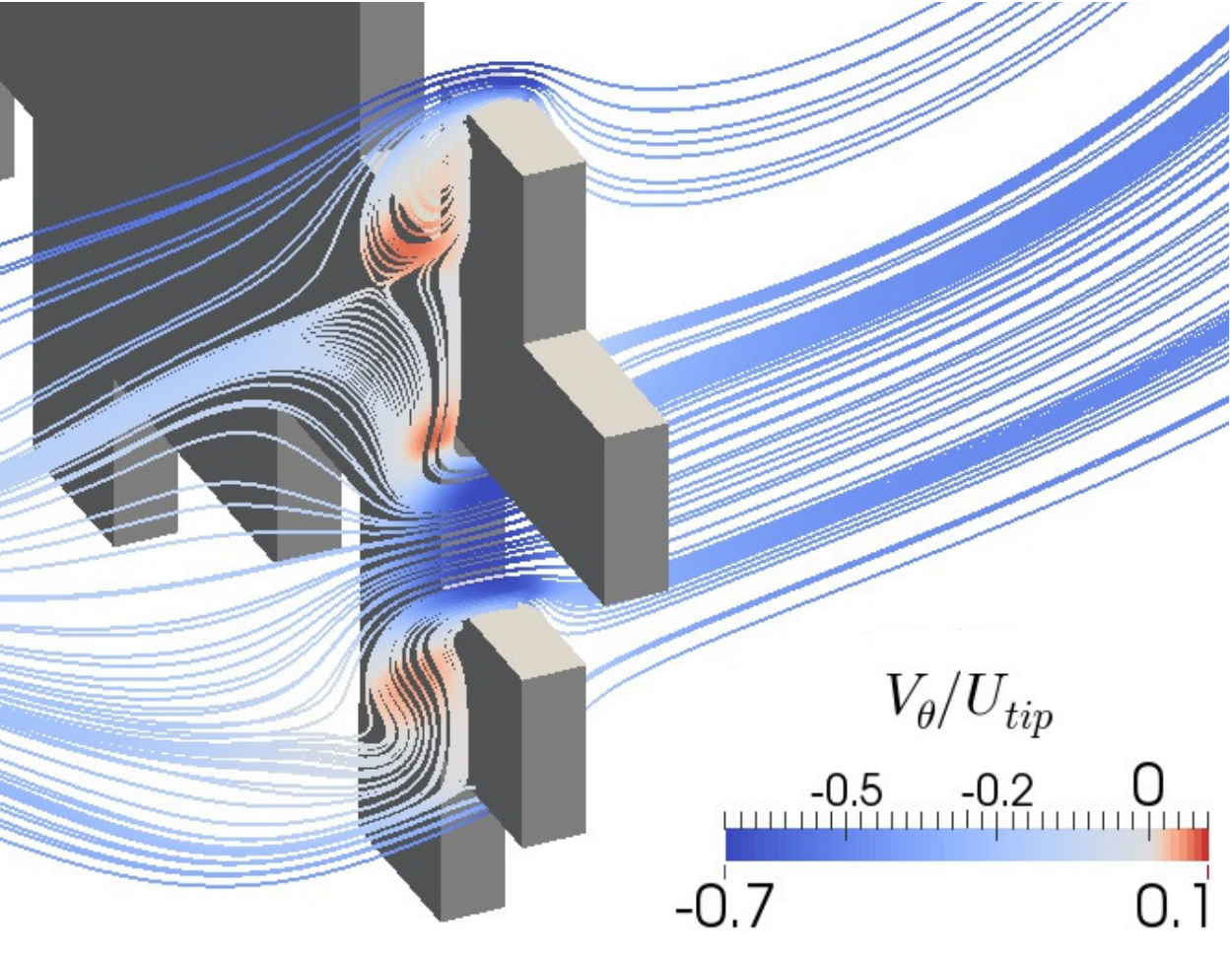}
        \caption{}
    \end{subfigure}  
  \caption{Recirculation zone behind the blades at $Re=1600$, (a)
    regular (b) fractal blade.}
  \label{fig:recirculation}
\end{figure}

In order to analyse further the mechanism that leads to a smaller
separation region in the suction side of the fractal blades, the
streamlines around both types of blades are illustrated in Figure
\ref{fig:recirculation}. To generate the streamlines, the radial
velocity component is subtracted from the mean flow field and only the
mean axial and azimuthal components are used. In other words, the
effect of radial jet is not illustrated in these figures. The
streamlines are coloured according to $V_{\theta}$. As seen in Figure
\ref{fig:recirculation} a), the flow passing along the upper and lower
edges of the regular blade recirculates in the wake and turns towards
the suction side.  With the influence of the radial jet along the
suction side of the blade, the recirculating fluid is carried towards
the blade tip, resulting in a spiraling motion that gives rise to the
trailing vortex pair.

In the case of the fractal impeller, the concave parts of the blade
perimeter allow the upcoming fluid with high azimuthal velocity, hence
high momentum, to enter into the recirculation region formed in the
wake. The jets penetrate and brake the two large recirculation zones,
leading to multiple smaller recirculation zones.

\begin{figure}[ht] 
  \centering   
    \begin{subfigure}[b]{0.552\textwidth}
    \centering   
        \includegraphics[width=\textwidth]{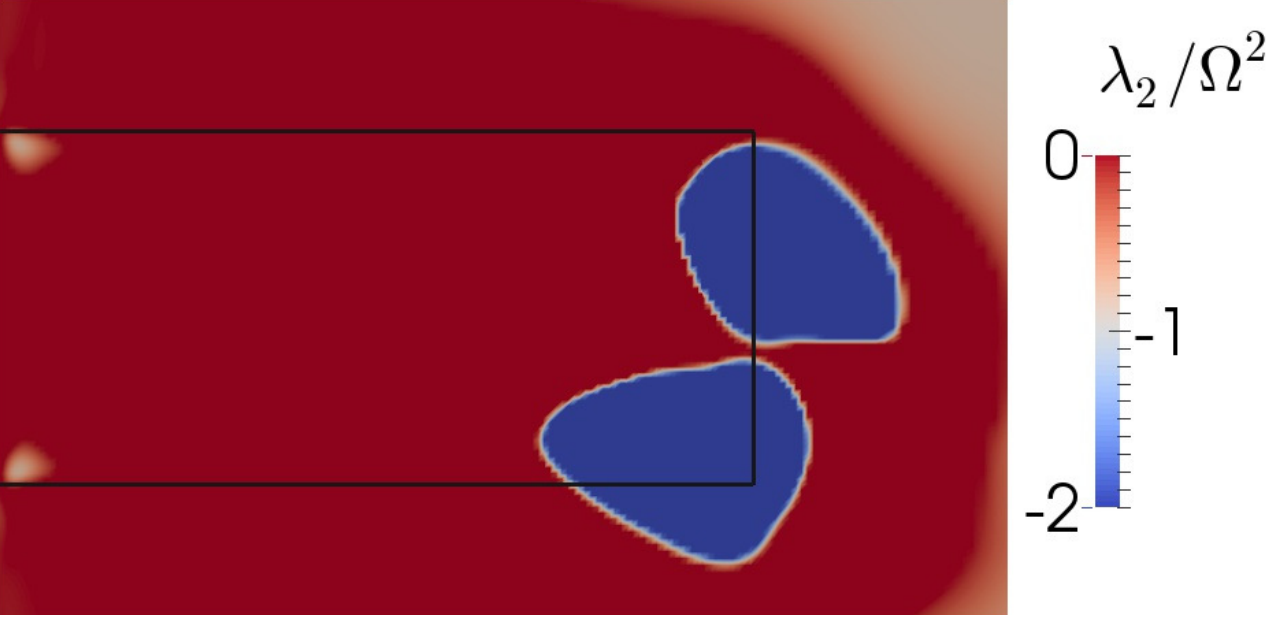}
        \caption{}
    \end{subfigure}  
    \begin{subfigure}[b]{0.432\textwidth}
    \centering 
        \includegraphics[width=\textwidth]{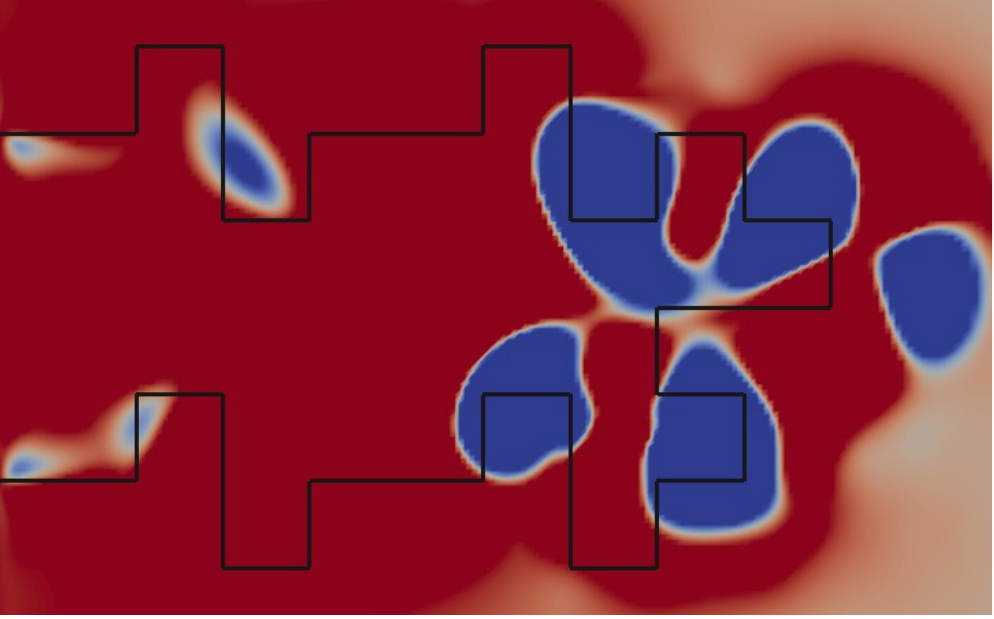}
        \caption{}
    \end{subfigure}  
  \caption{Vortex cores are illustrated using the contours of
    $\lambda_2/\Omega^2$ on a plane 15\textdegree\ behind the blades
    at $Re=1600$, (a) regular blade, (b) fractal blade. The blue regions
    in the figures display the vortex cores. Black lines indicate the
    borders of the impeller swept volume.}
  \label{fig:cores}
\end{figure}

This has a profound influence on the coherent trailing vortex
structures present in the wake of the blades. Vortex cores can be
visualised with the help of the $\lambda_2$-criterion introduced by
Jeong and Hussein.\cite{Jeong1995} In this criterion, $\lambda_2$ is
the second (intermediate) eigenvalue of the symmetric tensor
$\mathbf{S}^2+\mathbf{\Omega}^2$ (with $\lambda_3$ $\geq$ $\lambda_2$
$\geq$ $\lambda_1$), where $\mathbf{S}$ and $\mathbf{\Omega}$ are the
symmetric and antisymmetric parts of the velocity gradient tensor,
respectively. This is one of the widely used methods to illustrate
trailing vortices, employed also by Escudi\'e and
Lin\'e\cite{Escudie2007} and Sharp et al.\cite{Sharp2010} Figure
\ref{fig:cores} shows the contours of normalized $\lambda_2$ on a
plane 15\textdegree\ behind the blades, at $Re=1600$. The borders of
the impeller swept volume are marked with black lines and the blue
regions indicate the vortex cores. In the wake of the regular blade,
there are two large trailing vortex cores, as expected. On the other
hand, five separate cores appear in the wake of the fractal blade
(excluding the smaller ones closer to the shaft), which are about half
the size of the cores in the wake of the regular blade. Since the
recirculation region is penetrated by jets through the concave edges
of the fractal blade as shown in Figure \ref{fig:recirculation}b, the
coherent structures are broken into multiple trailing vortices. 

We can conclude that the pressure distribution on the suction side of
blades is determined by the trailing vortex structures that emerge in
the flow separated wake region. The wake properties are therefore
strongly affected by the blade design. This opens the possibility to
design blades with favourable characteristics that reduce the drag
coefficient and impeller torque, without changing the total area.

\vspace*{0.5cm}

\section*{Transport of angular momentum}
As a result of the above differences in the pressure distribution on
the blade surfaces, the regular impeller applies a higher torque to
the volume of fluid in the vessel compared to the fractal impeller at
$Re=1600$. Therefore, the flux of angular momentum away from the
impeller must be higher. Some questions naturally arise: Which
differences in the flow field around the impeller result in the higher
angular momentum flux? How is this related to the modification in the
blade shape or in the trailing vortex system? In this section we try
to answer these questions.

\begin{figure}[ht]
  \centering \includegraphics[width=0.35\textwidth]{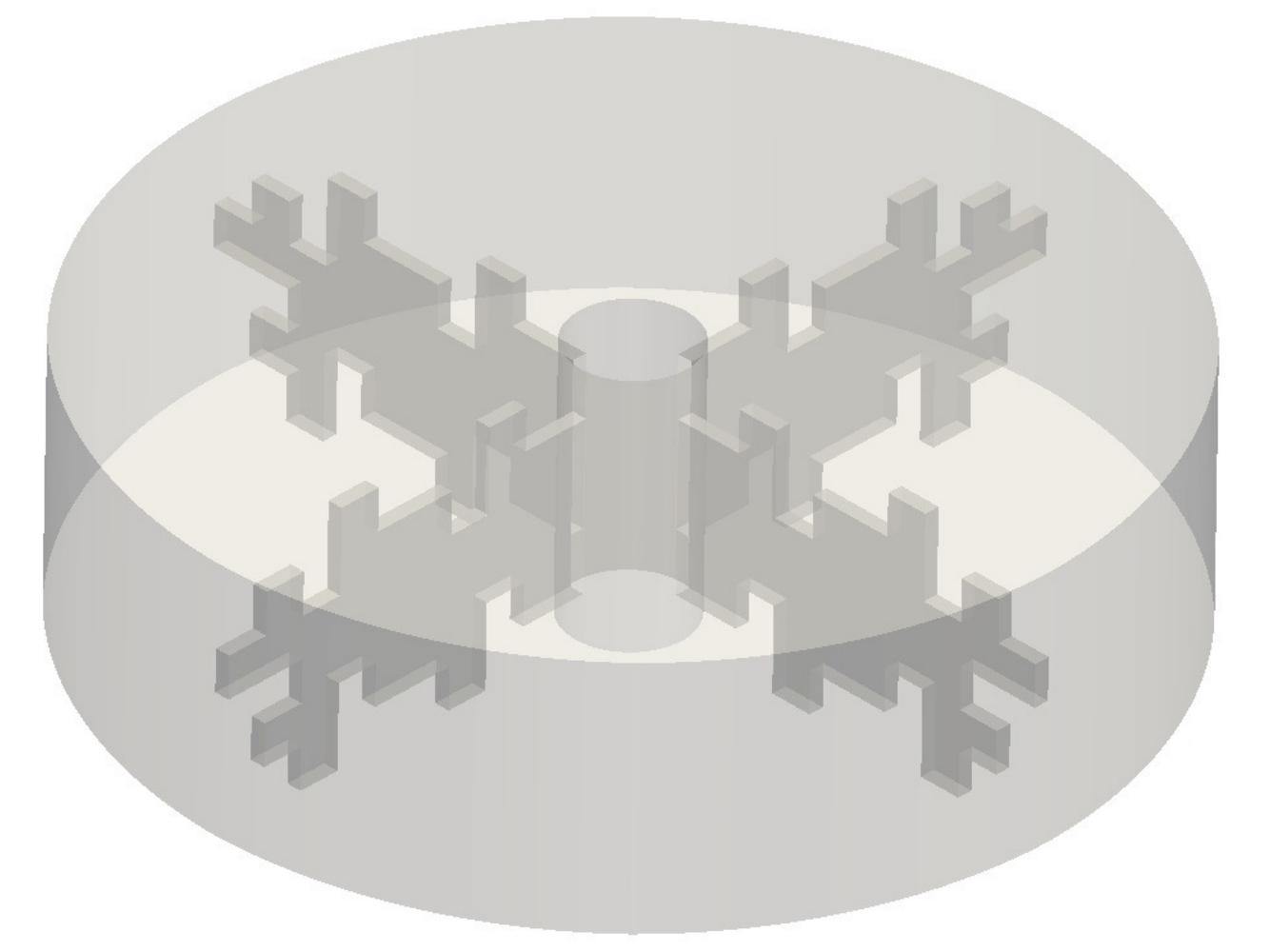}
  \caption{Control volume around the fractal impeller.}
  \label{fig:cv}
\end{figure}

To this end, we consider a control volume (CV) around the impeller as
shown in Figure \ref{fig:cv} and compute the angular momentum balance
in this CV. The impeller can be thought of as a source of angular
momentum which is then transported through the borders of the CV and
is eventually lost at the tank walls due to viscous
stresses.\cite{Basbug}

The general form of the angular momentum balance in this CV is:
\begin{equation}
T_{imp}= \left< \int_{S=S_{in}+S_{out}} (\vec{r} \times \vec{u})
\rho\vec{u}\cdot \: \vec{dS} - \int_{S=S_{in}+S_{out}} \vec{r} \times
(\boldsymbol\tau\cdot\vec{dS}) \right> \cdot\hat{e}_z ,
\label{eq:navier_ang}
\end{equation}
where $T_{imp}$ is the impeller torque, obtained from integration of
moment distribution due pressure and viscous forces over the impeller
blade (the major contributor is the pressure force). The angular
brackets $\left < \right >$ represent time-averaging in the inertial
frame, the first term in the right hand side is the net angular
momentum flux through the boundary $S$ of the CV, and the second term
is the moment due to viscous stresses on $S$. Vector $\hat{e}_z$ is
the unit vector along the impeller axis and $S_{in}$ is the part of
the CV's surface normal to $\hat{e}_z$ (both planar surfaces) whereas
$S_{out}$ is the remaining cylindrical part. Using the DNS data, both
sides of Equation \ref{eq:navier_ang} are computed separately and
balance up to three significant digits. Although the viscous term is
also taken into account during the evaluation of this balance, it
constitutes a small portion of the right-hand side. This term is about
$2-4\%$ of the total transport at $Re=320$ and this percentage is an
order of magnitude smaller at $Re=1600$.

Equation \ref{eq:navier_ang} is very general and it contains the effects of 
spatial variation of velocity as well as turbulence (due to time averaging
operation). Ignoring the viscous term and assuming that the velocity
is uniform and stationary, we get a simplified form, known as the
Euler's turbine equation\cite{White2011}
\begin{equation}
T_{imp}=\rho \, Q \, (L_{S_{out}}-L_{S_{in}}) \: ,
\label{eq:euler}
\end{equation}
where $Q$ is the volumetric flow rate through the CV (due to
continuity $Q_{S_{in}}=Q_{S_{out}}=Q$), while $L_{S_{in}}$ and
$L_{S_{out}}$ stand for the angular momentum per unit mass at the
inlet and outlet of the CV, respectively. The product
$\rho\,Q\,(L_{S_{out}}-L_{S_{in}})$ denotes the net angular momentum
flux through the boundary of the CV and it is the simplified form of
the first term on the right hand side of Equation \ref{eq:navier_ang}.

Equation \ref{eq:euler} indicates that, in order to examine in more
detail the differences between the standard and fractal impellers, it
is instructive to investigate the net flow rate through the CV. The
dimensions of the CV are selected such that it has the smallest
possible volume that can contain the fractal impeller, allowing a
small gap of $0.01\,R$ between edges of the blade and the CV borders,
hence $R_{CV}=1.11\,R$ (see Figure \ref{fig:cv}). For a fair
comparison, the CV used for the regular impeller has exactly the same
dimensions as the one used for the fractal impeller. The flow rate $Q$
can be computed by the integration of $U_{r}\, \hat{e}_r \cdot
\vec{n}$ over the side surface of the CV (hereafter $S_{out}$), where
$\vec{n}$ is the unity vector normal to the surface pointing outwards
from the CV (equal to $\hat{e}_r$ because the volume is
cylindrical). This gives the outflow from the CV, since strong radial
jets generated by the blades carry the fluid from the impeller towards
the walls. $Q$ can be also computed from integration of $U_{ax}\,
\hat{e}_z \cdot \vec{n}$ over the upper and lower flat surfaces of the
CV (hereafter $S_{in}$); that would give the inflow which is equal to
the outflow (with opposite sign).

\begin{table}[!ht] 
\caption{Normalized flow rate, $N_q=Q/(N D^3)$, through the CV in Figure \ref{fig:cv}. The difference between the flow rates of regular and fractal impellers is presented in the rightmost column.}
\begin{tabular}{ l || c | c | c } 
                & Regular    &  Fractal &  Difference   \\ \hline 
 $Re=320$       & $ 0.577 $  &  $ 0.621 $  & $ 7.3\% $    \\ 
 $Re=1600$      & $ 0.434 $  &  $ 0.481 $  & $ 10.3\% $   \\  
\end{tabular}
\label{table:flow_rate}
\end{table}

The values of the normalized flow rate, $N_q=Q/(N D^3)$, are reported
in Table \ref{table:flow_rate} for the four cases examined. It is
noted that $N_q$ decreases with $Re$, which is in agreement with the
trend presented by Nagata\cite{Nagata1975} for an unbaffled tank with
an eight-bladed paddle impeller. He showed that $N_q$ increases
rapidly in the laminar regime and reaches a peak at $Re\approx90$,
where $N_q\approx1$. This is followed by a slow decrease over
transitional and turbulent regimes down to $N_q\approx0.45$ at
$Re=10^6$. Since we employ a four-bladed impeller, the values are
expected to be somewhat lower compared to those of Nagata, hence the
quantities presented in Table \ref{table:flow_rate} are reasonable.

On the other hand, it may be surprising that the fractal impeller has
at $Re=320$ ca. $7\%$ higher $N_q$ compared to the regular impeller
with equal $N_p$, and at $Re=1600$ ca. $10\%$ higher $N_q$ despite
having $8\%$ lower $N_p$.  This difference in $N_q$ in favour of the
fractal impeller is even larger at $Re=1.5\times 10^5$, according to
the PIV study of Steiros et al.\cite{Steiros2017} This property of the
fractal impeller has the potential to accelerate the stirring of a
scalar injected into the fluid, which is especially desirable for low
to moderate $Re$, where the macro-mixing plays an important role. The
study of the mixing properties of fractal impellers will be the
subject of future research.

\begin{figure}[!ht]
  \centering   
  \includegraphics[width=0.6\textwidth]{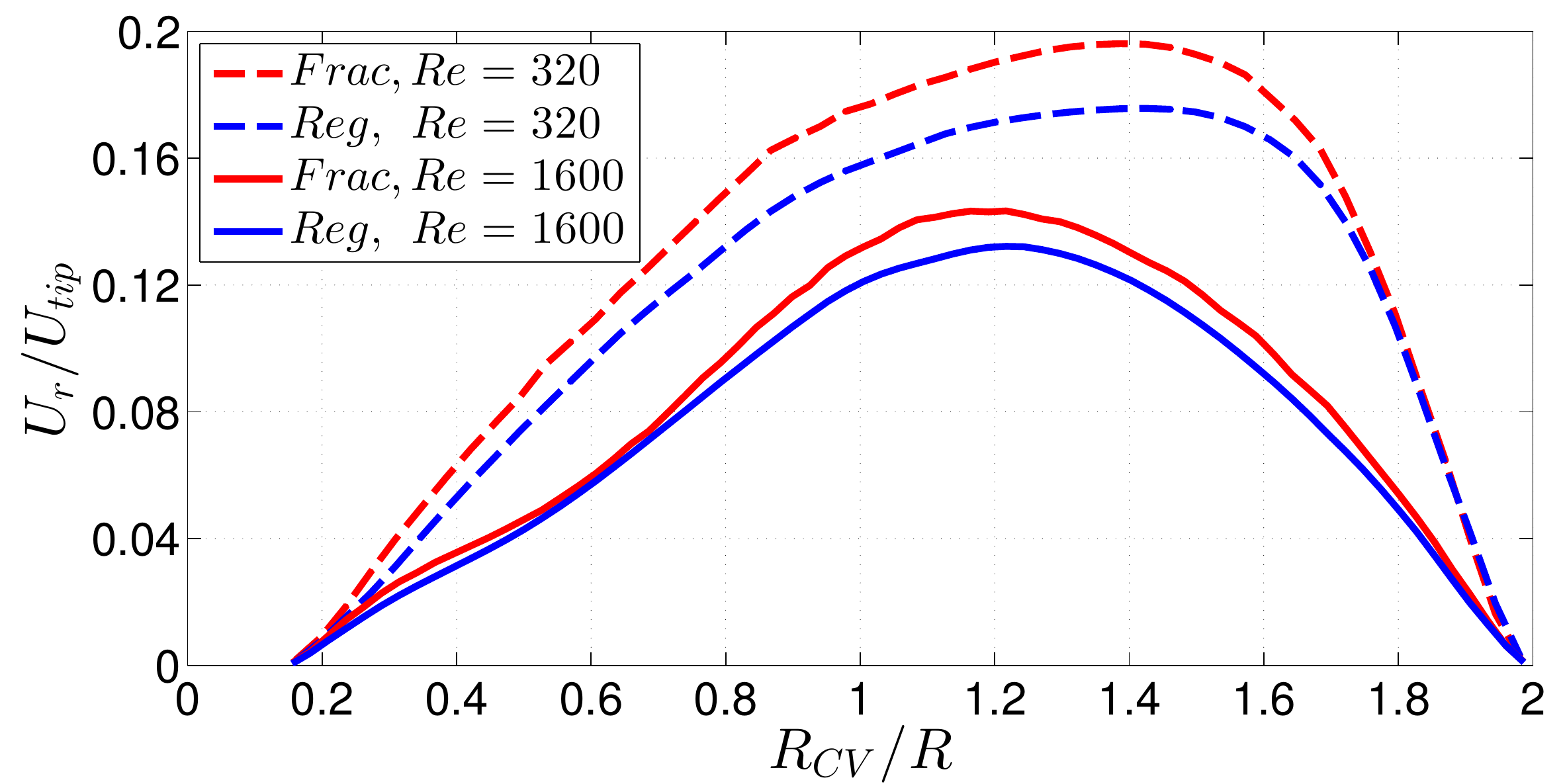} 
  \caption{Radial velocity averaged over $S_{out}$ and in time, for varying $R_{CV}$.}
  \label{fig:urad}
\end{figure}

To evaluate the dependence of this result on the size of the CV, the
mean radial velocity averaged over $S_{out}$ for varying CV-radius is
plotted in Figure \ref{fig:urad}. Since this calculation corresponds
to averaging $U_r$ in the azimuthal direction, it is the same in both
absolute and relative reference frames. The radial flow rate is
consistently higher for the fractal impeller over a broad range of CV
radii $(1<R_{CV}/R<1.6)$. It is deduced therefore that the higher power
number of the regular impeller at $Re=1600$ is not related to an
increased flow rate, but to a larger difference of the fluid angular
momentum per unit mass between $S_{in}$ and $S_{out}$. Therefore, the
crucial quantity to evaluate is the transport of angular momentum into
and out of the CV per unit flow rate.

The aforementioned transport occurs via two mechanisms: advective
transport and viscous transport due to the shear stresses (refer to
Equation \ref{eq:navier_ang}), but the latter has a negligible contribution. 
The advective transport of angular momentum over $S_{in}$ and $S_{out}$ 
is normalized with the mass flow rate $(\rho\,Q)$ and made nondimensional 
using the angular momentum (per unit mass) at the blade tip $(R\,U_{tip})$, as follows:

\begin{equation}
l_{in} = \frac{|\int_{S_{in}} \rho r \left< u_{\theta} u_z \right> \:\hat{e}_z \cdot d\vec{S}|}{\rho\,Q\,R\,U_{tip}},\:\:\:\:\:
l_{out}= \frac{\int_{S_{out}} \rho r \left<  u_{\theta} u_r \right> \:\hat{e}_r \cdot d\vec{S}}{\rho\,Q\,R\,U_{tip}}.  
\label{eq:l_in_out}
\end{equation}
We define $l_{in}$ in terms of the absolute value of the integral in Equation
\ref{eq:l_in_out}, since this integral results in a negative value
indicating momentum entering into the CV. Therefore the net normalized
advective transport is $\Delta l=l_{out}-l_{in}$. The values of
$l_{in}$, $l_{out}$ and $\Delta l$ are listed in Table
\ref{table:angflux} for both $Re$ and impeller types.

\begin{table}[!ht] 
\caption{The values of $l_{in}$ and $l_{out}$ as defined by Equation \ref{eq:l_in_out}. The values of $\Delta l$ obtained with regular and fractal impellers are compared and the percentage difference is also shown.}
\begin{tabular}{ l l || c | c | c } 
         &      &$l_{in}$ &$l_{out}$ &$\Delta l=l_{out}-l_{in}$  \\ \hline 
$Re=320$ & Reg. &$0.159 $ &$0.557 $  &$0.398 $                   \\ 
         & Frac.&$0.177 $ &$0.536 $  &$0.359 $                   \\ 
         &      &         &          &$=>10\%$ difference        \\ \hline
$Re=1600$& Reg. &$0.288 $ &$0.656 $  &$0.368 $                   \\  
         & Frac.&$0.283 $ &$0.589 $  &$0.306 $                   \\  
         &      &         &          &$=>18\%$ difference        \\ \hline         
\end{tabular}
\label{table:angflux}
\end{table}

When we compare the two values of $Re$, we see that $l_{in}$ is
significantly higher at $Re=1600$, meaning that the fluid enters the
CV with a higher angular momentum. This can be explained as follows:
While the radially discharged flow circulates back to the impeller, it
loses part of its angular momentum due to shear stresses applied by
tank walls. This effect of the wall shear stress weakens with growing
$Re$, leading to a higher $l_{in}$. If there were baffles on the tank
walls, $l_{in}$ would be much smaller, since viscous forces on the
walls are strongly augmented by pressure forces on the baffle surfaces
normal to the azimuthal fluid motion. This would lead to a much larger
$\Delta l$, hence larger impeller torque and $N_p$. Moreover, Table
\ref{table:angflux} shows that $l_{out}$ also increases with $Re$, but
not as much as $l_{in}$, resulting in a lower $\Delta l$ at $Re=1600$
for both types of impellers. Consequently, the computed $N_p$
decreases with $Re$.

At both Reynolds numbers, the regular impeller yields a higher $\Delta
l$ with respect to the fractal impeller. However, at $Re=320$ this
difference is compensated mainly by the higher flow rate of the
fractal impeller, and also with the inclusion of the small
contribution of the viscous transport over $S_{out}$ (since CV-borders
are very close to the tip of the fractal blade, the mean velocity
gradients are high at $S_{out}$ near the blade tip). Taking everything
into account, at the low Re cases both types of impeller have the same
torque and $N_p$. On the other hand at $Re=1600$, the $18\%$ higher
$\Delta l$ of the regular impeller is only partially balanced by the
$10\%$ lower flow rate. Consequently, the regular impeller requires
$8\%$ higher power consumption compared to the fractal impeller.

\begin{figure}[ht!]
  \centering   
    \begin{subfigure}[c]{0.49\textwidth}
    \centering   
        \includegraphics[width=\textwidth]{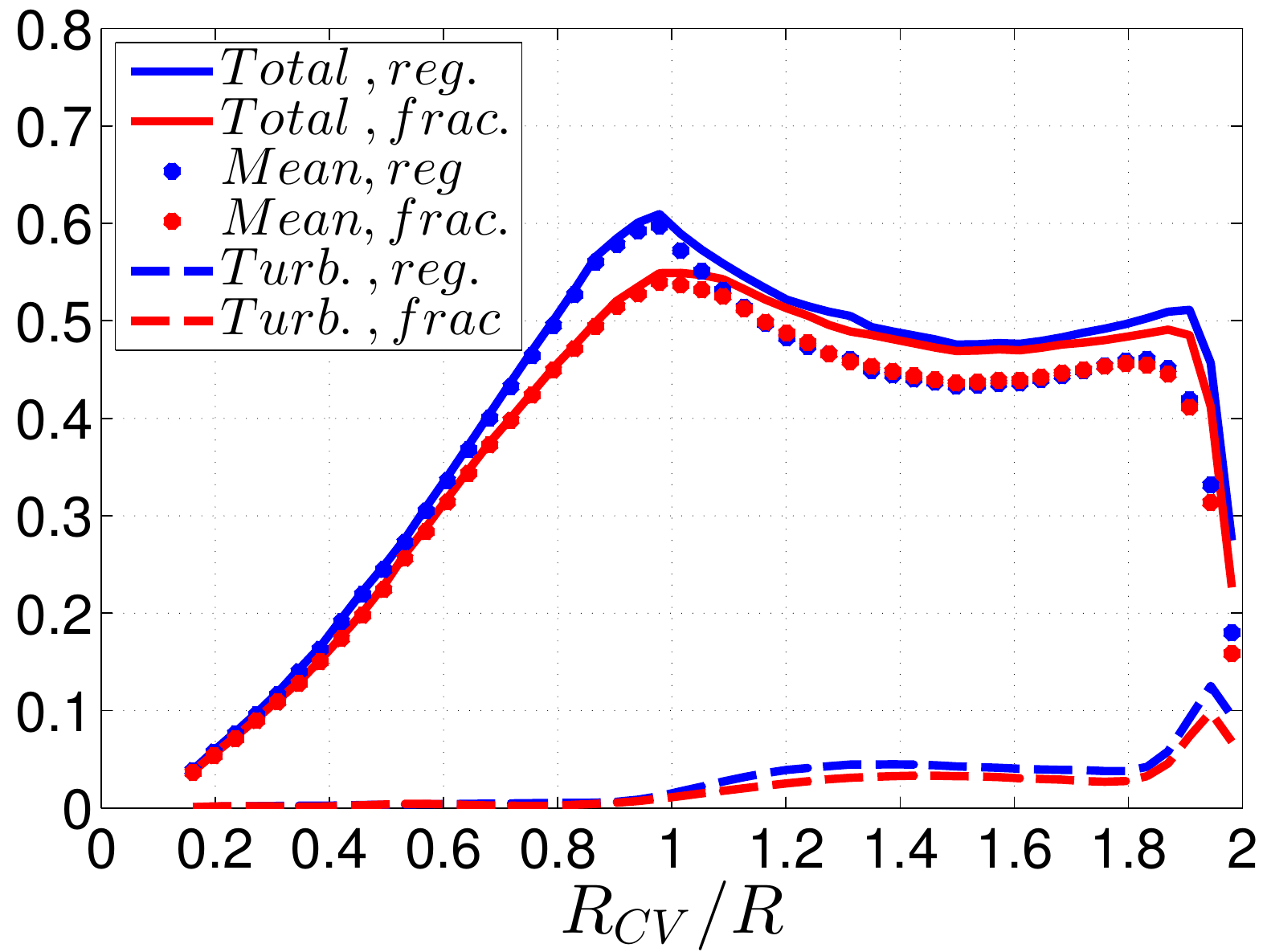}
        \caption{}
    \end{subfigure}  
    \begin{subfigure}[c]{0.49\textwidth}
    \centering 
        \includegraphics[width=\textwidth]{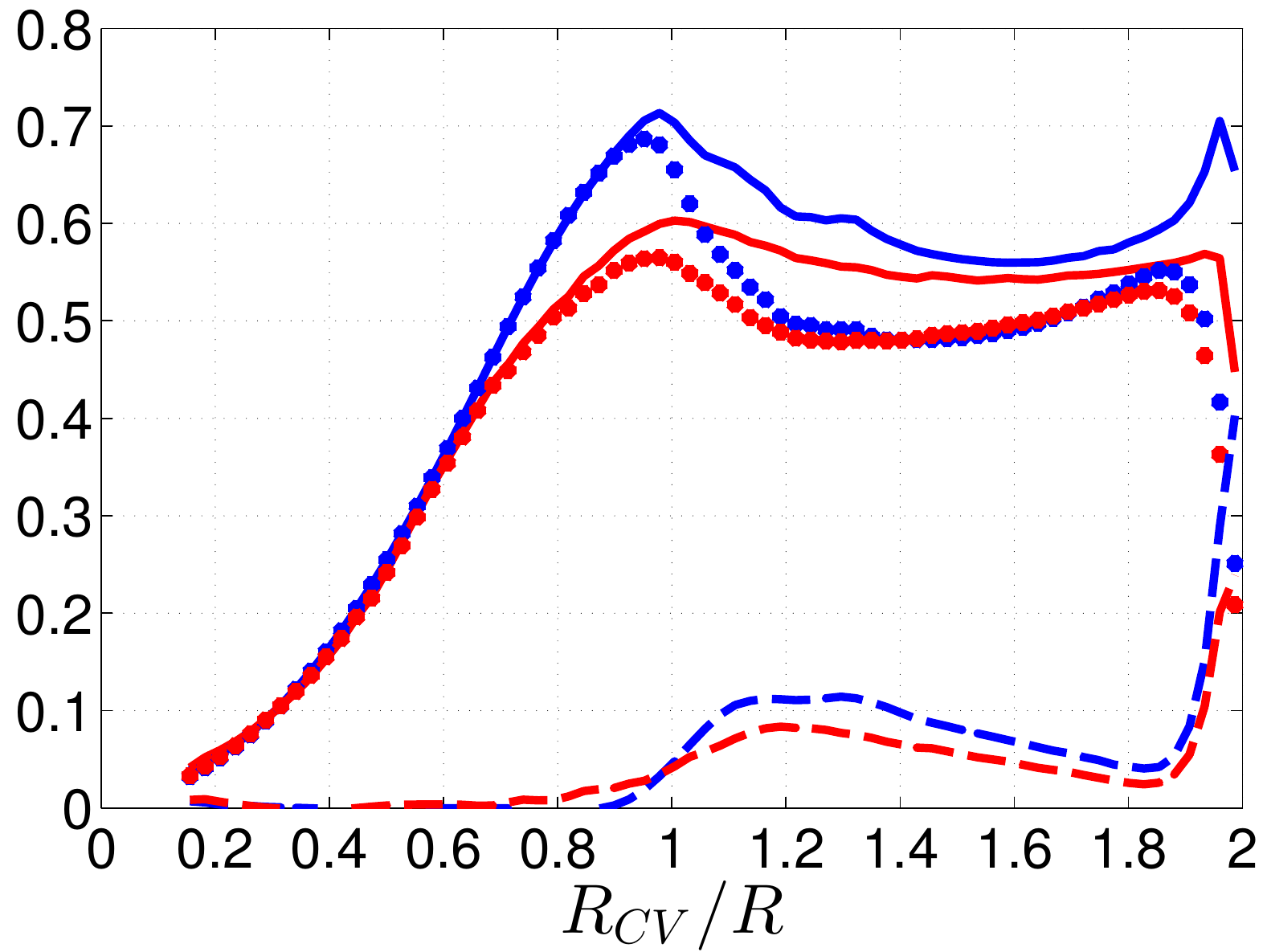}
        \caption{}
    \end{subfigure}  
  \caption{$l_{out}$ (Total), $l_{out,mean}$ (Mean) and $l_{out,turb}$
    (Turb.) computed at $S_{out}$ with varying radial positions ($R_{CV}$) for
    regular and fractal impellers (a) at $Re=320$, (b) at $Re=1600$.}
  \label{fig:l_out}
\end{figure}

In order to provide more insight as to why $\Delta l$ is larger for
the regular impeller and why this difference grows with $Re$, the
advective transport term is decomposed in two parts, representing the
contributions of the mean and fluctuating velocities. So the term
$r\left< u_{\theta} u_r \right>$ is decomposed as follows:
\begin{equation}
r\left< u_{\theta} u_r \right>= r U_{\theta} U_r + r\left< u'_{\theta} u'_r \right>.
\label{eq:decompose}
\end{equation}
Using this expression, we can evaluate separately $l_{out,mean}$,
which is the contribution due to mean velocities (i.e. $r U_{\theta}
U_r$), and $l_{out,turb}$, the contribution due to turbulent
fluctuations (i.e. $r\left< u'_{\theta} u'_r \right>$).
\begin{equation}
l_{out,mean}= \frac{\int_{S_{out}} r U_{\theta} U_r \:\hat{e}_r \cdot
  d\vec{S}}{\,Q\,R\,U_{tip}},\:\:\:\:\: l_{out,turb}=
\frac{\int_{S_{out}} r \left< u'_{\theta} u'_r \right> \:\hat{e}_r
  \cdot d\vec{S}}{\,Q\,R\,U_{tip}} .
\label{eq:l_out_decomp}
\end{equation}
Both terms are separately integrated over the surface $S_{out}$ of the
CV and normalized as in Equation \ref{eq:l_out_decomp}. The radial
profiles of $l_{out,mean}$ and $l_{out,turb}$ are illustrated in
Figure \ref{fig:l_out}. A similar decomposition is also performed for
$l_{in}$ to evaluate $l_{in,mean}$ and $l_{in,turb}$ separately, but
it is not discussed here because the contribution of $l_{in,turb}$ is
found to be ca. 1\% or less.

\begin{figure}[ht!]
  \centering   
    \begin{subfigure}[c]{0.49\textwidth}
    \centering   
        \includegraphics[width=\textwidth]{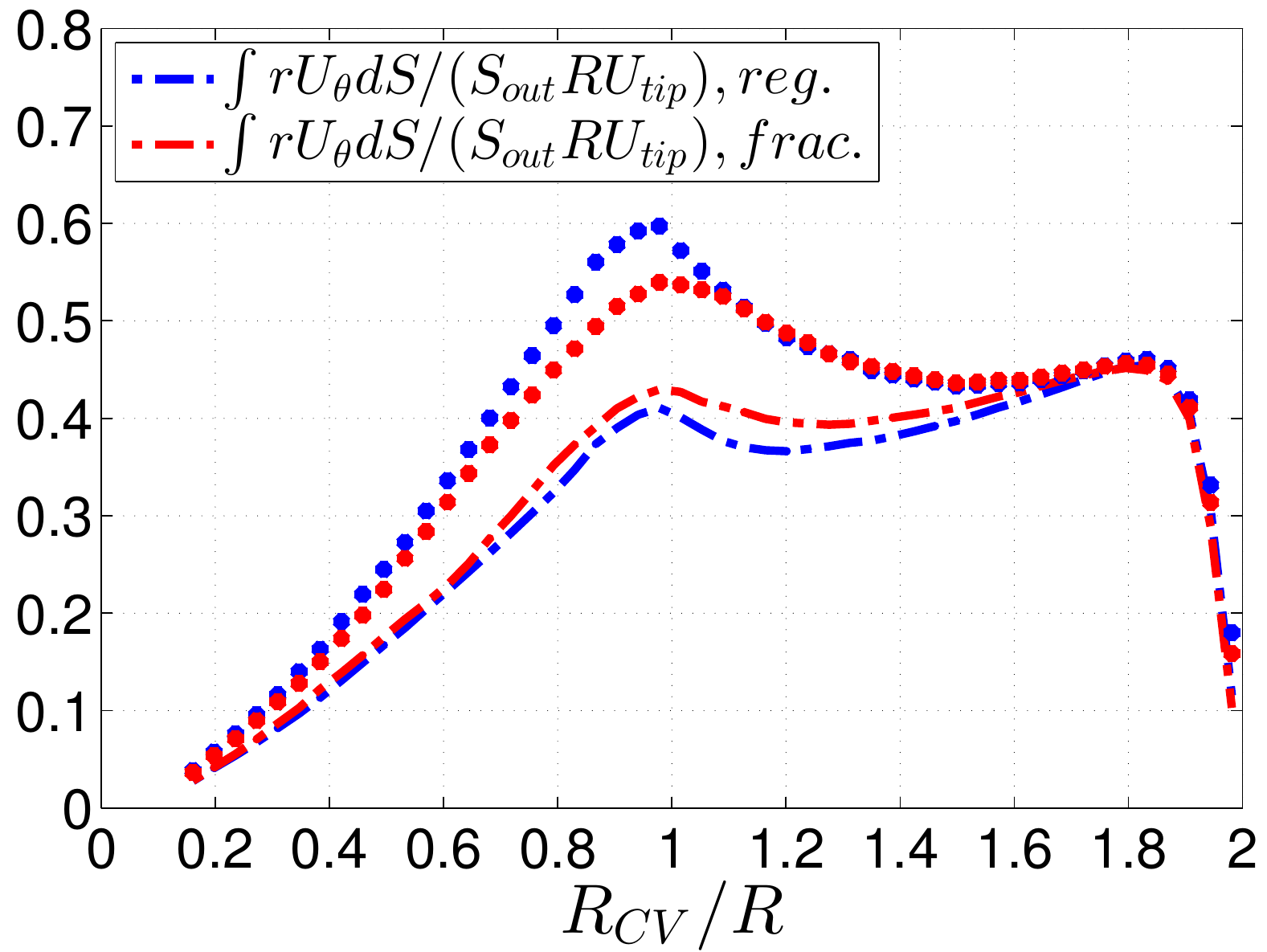}
        \caption{}
    \end{subfigure}  
    \begin{subfigure}[c]{0.49\textwidth}
    \centering 
        \includegraphics[width=\textwidth]{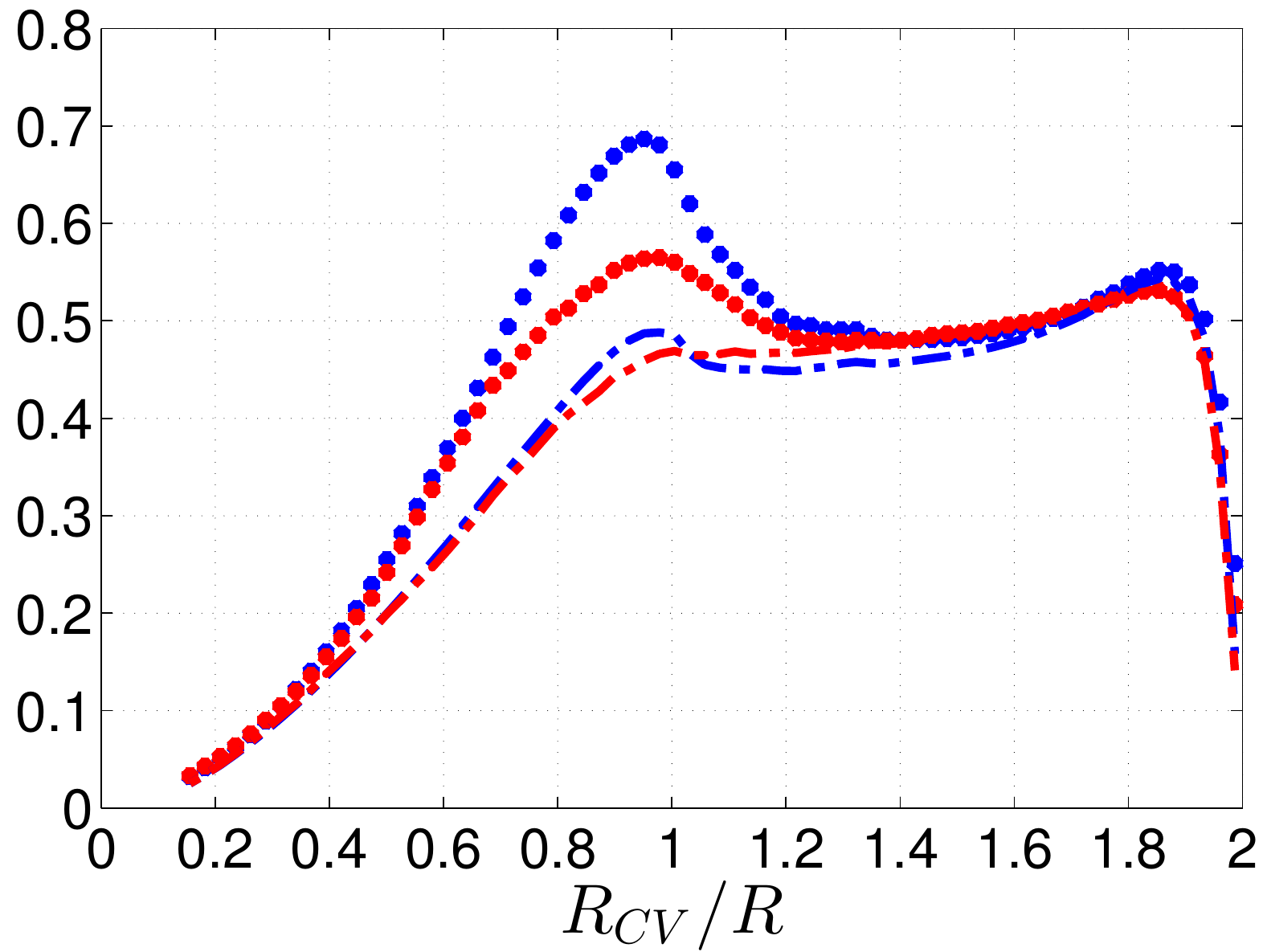}
        \caption{}
    \end{subfigure}  
  \caption{$l_{out,mean}$ shown with dots as in Figure \ref{fig:l_out}
    and compared with mean angular momentum ($r U_{\theta}$) averaged
    over $S_{out}$ with varying radial positions ($R_{CV}$) for regular and
    fractal impellers that is shown with dash-dotted lines (a) at
    $Re=320$, (b) at $Re=1600$.}
  \label{fig:l_mean}
\end{figure}

Figure \ref{fig:l_out} shows that for both impeller types and at both
$Re$, $l_{out,mean}$ is larger than $l_{out,turb}$, therefore it
determines the form of the $l_{out}$ profile. Since $l_{out,mean}$ has
in the denominator the flow rate for normalization, which is basically
the integral of $U_r$ over $S_{out}$, it might have been assumed that
the profile of $l_{out,mean}$ corresponds to the profile of angular
momentum $(r U_{\theta})$ averaged over $S_{out}$. This would be
correct if $U_r$ and $U_{\theta}$ had a homogeneous spatial
distribution over $S_{out}$. To see if this is the case, angular
momentum $(r U_{\theta})$ is averaged over $S_{out}$ with varying
radius ($R_{CV}$) and is normalized with $R U_{tip}$. The so-obtained radial
profile of angular momentum is illustrated in Figure \ref{fig:l_mean}
with dash-dotted lines and is compared with the profile of
$l_{out,mean}$. It is seen in Figures \ref{fig:l_mean} a) and b), that
the angular momentum increases monotonically from the shaft until the
blade tip (for $R_{CV}/R \leqslant 1$) as in the core of a forced vortex,
then remains approximately constant until the near-wall region (for
$1<R_{CV}/R<1.9$) which is expected in a free vortex. On the other hand,
the curves of $l_{out,mean}$ have a higher slope until $R_{CV}/R \approx
1$, where they reach their maxima and have a significant surplus
compared to the average angular momentum, for both impeller types and
$Re$ numbers. This surplus is due to the strong spatial correlation
between $U_r$ and $U_{\theta}$ in the impeller discharge
region. Especially on the suction side of the blades near the blade
tip $(0.8<R_{CV}/R<1)$, we observe the highest radial and azimuthal
velocities approximately in the same region. This spatial correlation
is stronger for the regular impeller, hence it has a higher
$l_{out,mean}$, despite the similar angular momentum profiles of both
impeller types. But the difference between the regular and fractal
impellers is observed only until a certain radial position, $R_{CV}/R
\approx 1.1-1.2$, where the profiles of $l_{out,mean}$ of both
impellers merge. At $Re=1600$, the profiles of $l_{out,mean}$ also
collapse with the profiles of angular momentum at $R_{CV}/R \approx
1.2$. This indicates that the spatial distributions of $U_r$ and
$U_{\theta}$ over $S_{out}$ are homogenized. This happens
approximately at the radial position where $l_{out,turb}$ completes
its growth and reaches a plateau (see Figure \ref{fig:l_out}b). It can
be deduced that the strong fluctuations (i.e. $\left< u'_{\theta} u'_r
\right>$) enhance the homogenization of mean velocities. However, at
$Re=320$ this homogenization occurs later, at $R_{CV}/R \approx 1.6$,
possibly due to the lack of turbulent mixing of momentum.

Focusing on Figure \ref{fig:l_out} and comparing the results at both
$Re$ numbers, several aspects can be clarified. At $Re=320$, the
difference between the $l_{out}\,$-profiles of regular and fractal
impellers is very small (except for the radial range of $0.7<R_{CV}/R<1.1$,
that is mainly due to $l_{out,mean}$). At $Re=1600$, on the other
hand, the difference between the $l_{out}\,$-profiles is larger for
$R_{CV}/R>0.6$. Since $l_{out,mean}\,$-profiles of regular and fractal
impellers collapse at $R_{CV}/R \approx 1.2$, any difference in the
transport mechanism is due to $l_{out,turb}$ for $R_{CV}/R>1.2$. Most
importantly, $l_{out,turb}$ grows almost threefold when $Re$ is
increased from $320$ to $1600$. Actually at both $Re$, regular
impeller yields ca. $30\%$ higher $l_{out,turb}$ than the fractal
impeller for the range of radial positions $1.1<R_{CV}/R<1.6$, but at
$Re=320$ the influence of $l_{out,turb}$ is not crucial. In
conclusion, at $Re=1600$, $l_{out,turb}$ accounts for the difference
in the transport of the angular momentum away from the impeller when
the regular and fractal impellers are compared. The larger shaft
torque of regular impeller, hence the larger source of angular
momentum inside the CV, is balanced in this way with the larger
transport over $S_{out}$.

The experiments of Steiros et. al \cite{Steiros} demonstrated
$11-12\%$ higher torque for the regular impeller compared to the
fractal impeller at $Re=1-2\times10^5$. Therefore, it can be expected
that $\left< u'_{\theta} u'_r \right>$ grows further with $Re$, hence
the contribution of $l_{out,turb}$ grows as well, until a fully
turbulent regime is reached.

\begin{figure}[ht] 
  \centering   
    \begin{subfigure}[b]{0.49\textwidth}
    \centering   
        \includegraphics[width=0.9\textwidth,trim={0cm 0cm 0cm 0.25cm},clip]{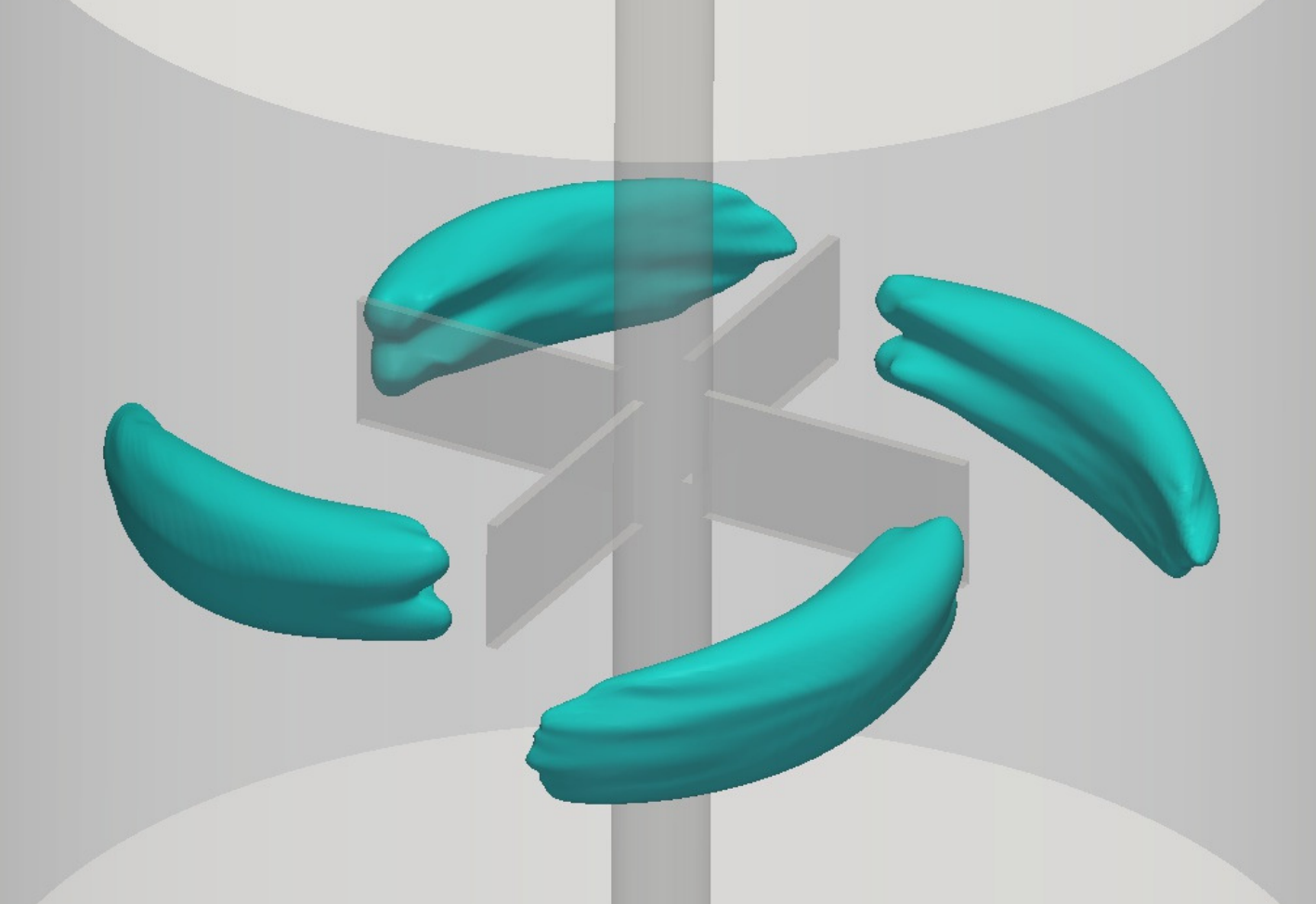}
       \caption{}
    \end{subfigure}  
    \begin{subfigure}[b]{0.49\textwidth}
    \centering 
        \includegraphics[width=0.9\textwidth]{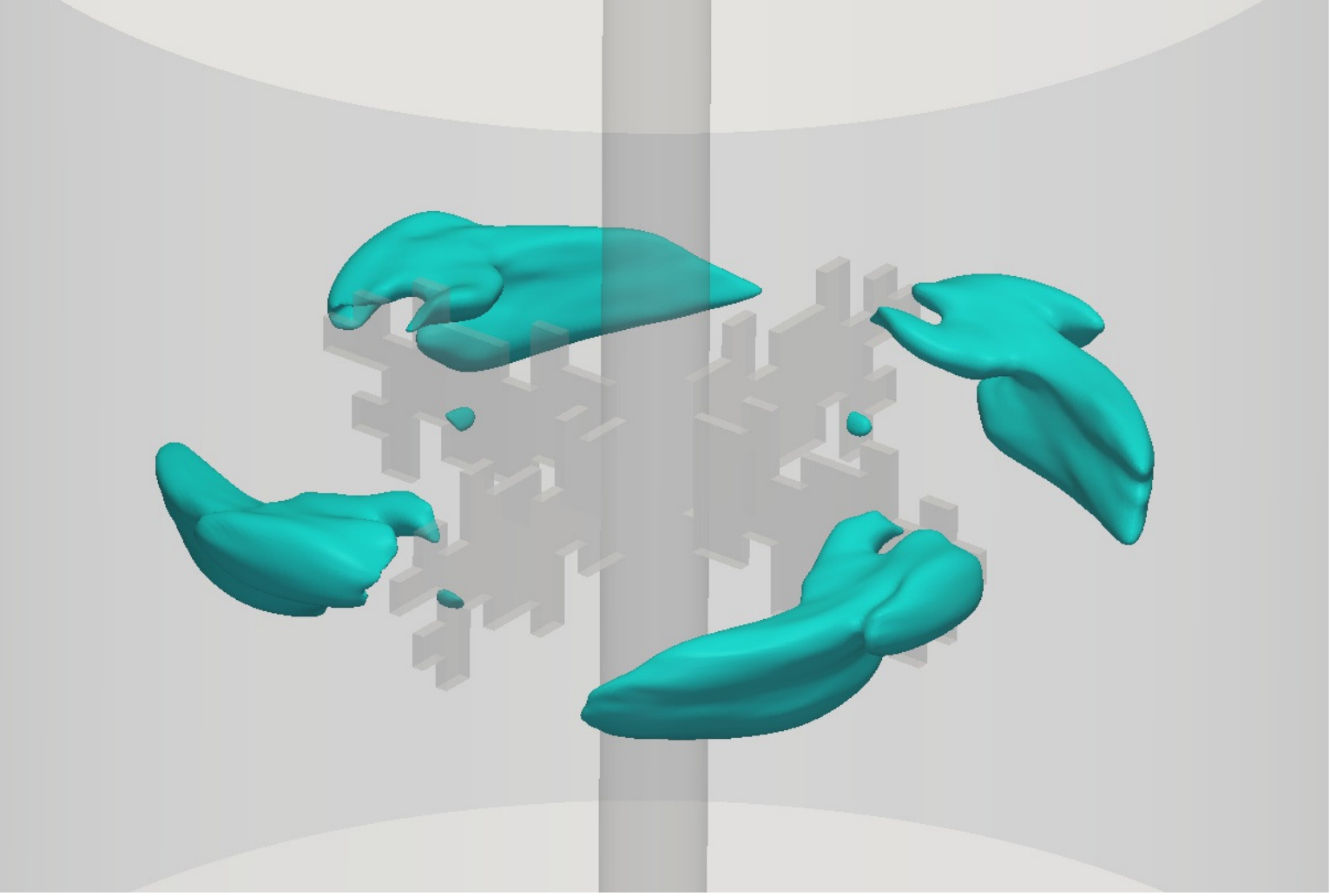}
        \caption{}
    \end{subfigure}  
  \caption{The region in the flow field where $r \left< u'_{\theta} u'_r \right>/(R U_{tip}^2) >2\%$ at $Re=1600$ for (a) regular, (b) fractal impeller.}
  \label{fig:max_trans}
\end{figure}

In addition to the radial range discussed above, where $l_{out,turb}$
plays an important role, the exact location can be determined in the
flow field, where the value of $r\left<u'_{\theta} u'_r \right>$
reaches its maximum level. This is illustrated in Figure
\ref{fig:max_trans} with the isosurfaces of $r \left< u'_{\theta} u'_r
\right>/(R U_{tip}^2)=2\% $, at $Re=1600$ for both impeller types. In
the volume contained in these isosurfaces, the value of $r \left<
u'_{\theta} u'_r \right>/(R U_{tip}^2)$ reaches a maximum of $14\%$
for the regular impeller and $6\%$ for the fractal impeller. It is
remarkable that the regions indicated by these isosurfaces coincide
exactly with the location of trailing vortices in the wake of the
blades. Therefore it is deduced that these coherent structures account
for the turbulent transport of the angular momentum radially away from
the impeller, quantified by $l_{out,turb}$.

The two large trailing vortices observed in the wake of the regular
blades are replaced with multiple and weaker vortices in the wake of
the fractal blade, as shown earlier in Figure \ref{fig:cores}. Still,
the integral of turbulent kinetic energy, i.e. $0.5 \left< u'_i u'_i
\right>$, over the entire volume is equal for both impellers at
$Re=1600$, but it is rather distributed in the fractal impeller
case. On the other hand, the correlation between the radial and
azimuthal velocity fluctuations, hence $l_{out,turb}$, is lower around
the fractal impeller, due to the break-down of trailing vortices.

It was previously shown that the larger recirculation region in the
wake of the regular blade accounts for the lower pressure on the
suction side, hence the higher form drag. With the help of Figure
\ref{fig:recirculation} it was also illustrated that this
recirculation leads to the formation of the trailing
vortices. Finally, it is shown that these coherent structures have a
strong influence on the transport of angular momentum. Taking these
into account, we can conclude that the differences in the form drag,
impeller torque and transport of the angular momentum between the
regular and fractal impellers are directly connected via the
modification of the trailing vortices.

\section*{Energy dissipation characteristics}

The time-average of the power draw of the impeller must be balanced by
the time-average of the total energy dissipation over the entire
tank. Indeed, the total dissipation in the tank is 8\% lower when the
fractal impeller is employed instead of the regular impeller (see
Table \ref{table:all_power}). In the previous section we analysed the 
mechanism leading to the differences in impeller torque and angular 
momentum transport. In this section we untangle how, 
and in which part of the flow field, the dissipation of
kinetic energy differs between the two impellers. We focus mainly on
$Re=1600$ and we consider time-average quantities. We stress the
time-average aspect of this section's study because, as we show in the
Appendix, both the power and the total energy dissipation fluctuate in
time and do not balance instantaneously as there is a cascade time-lag
between them.

\begin{figure}[ht] 
  \centering   
    \begin{subfigure}[b]{0.49\textwidth}
    \centering   
        \includegraphics[width=\textwidth]{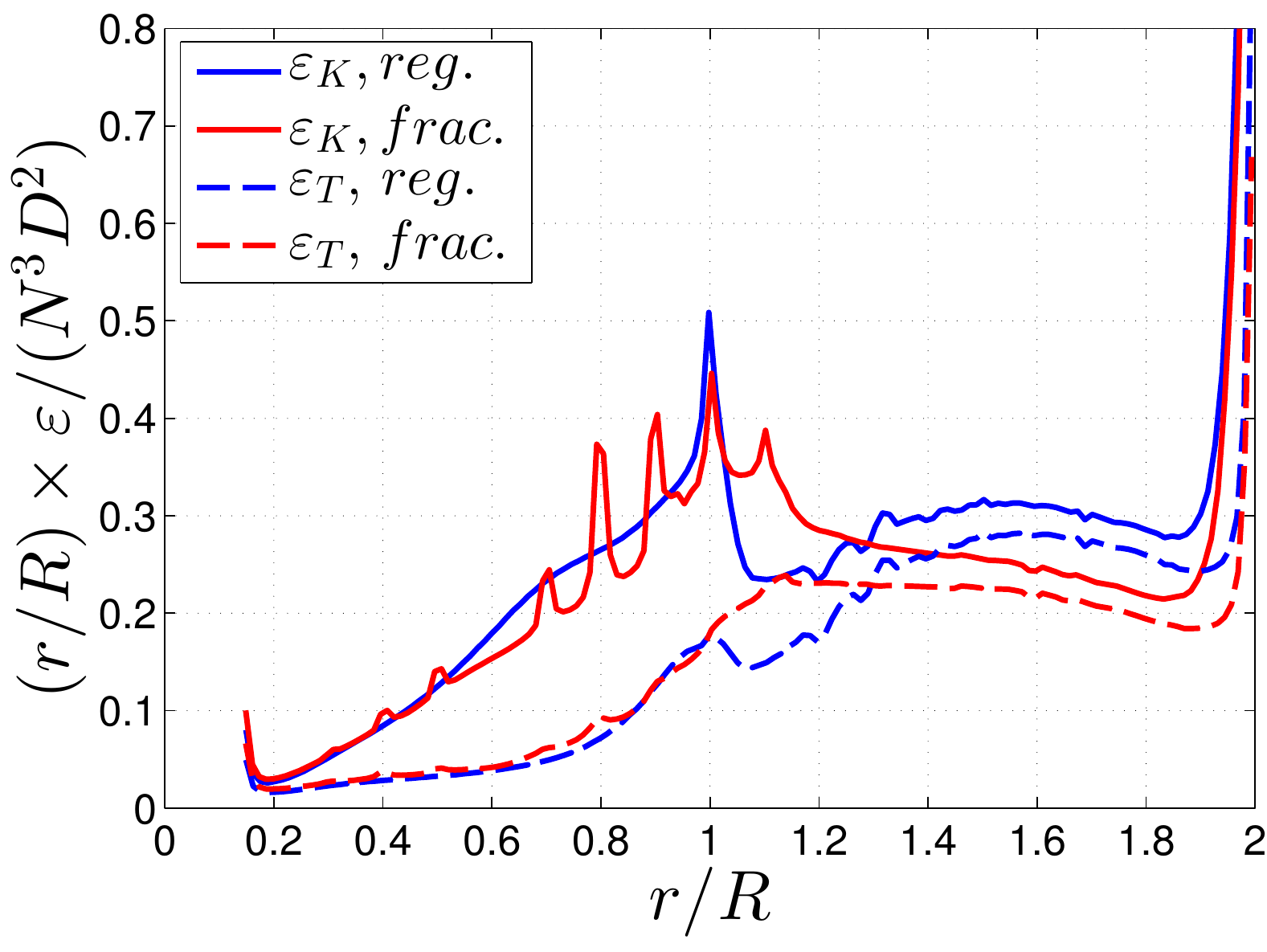}
        \caption{$-1/2<z/H<1/2$ (entire tank)}
    \end{subfigure}  
    \begin{subfigure}[b]{0.49\textwidth}
    \centering 
        \includegraphics[width=\textwidth]{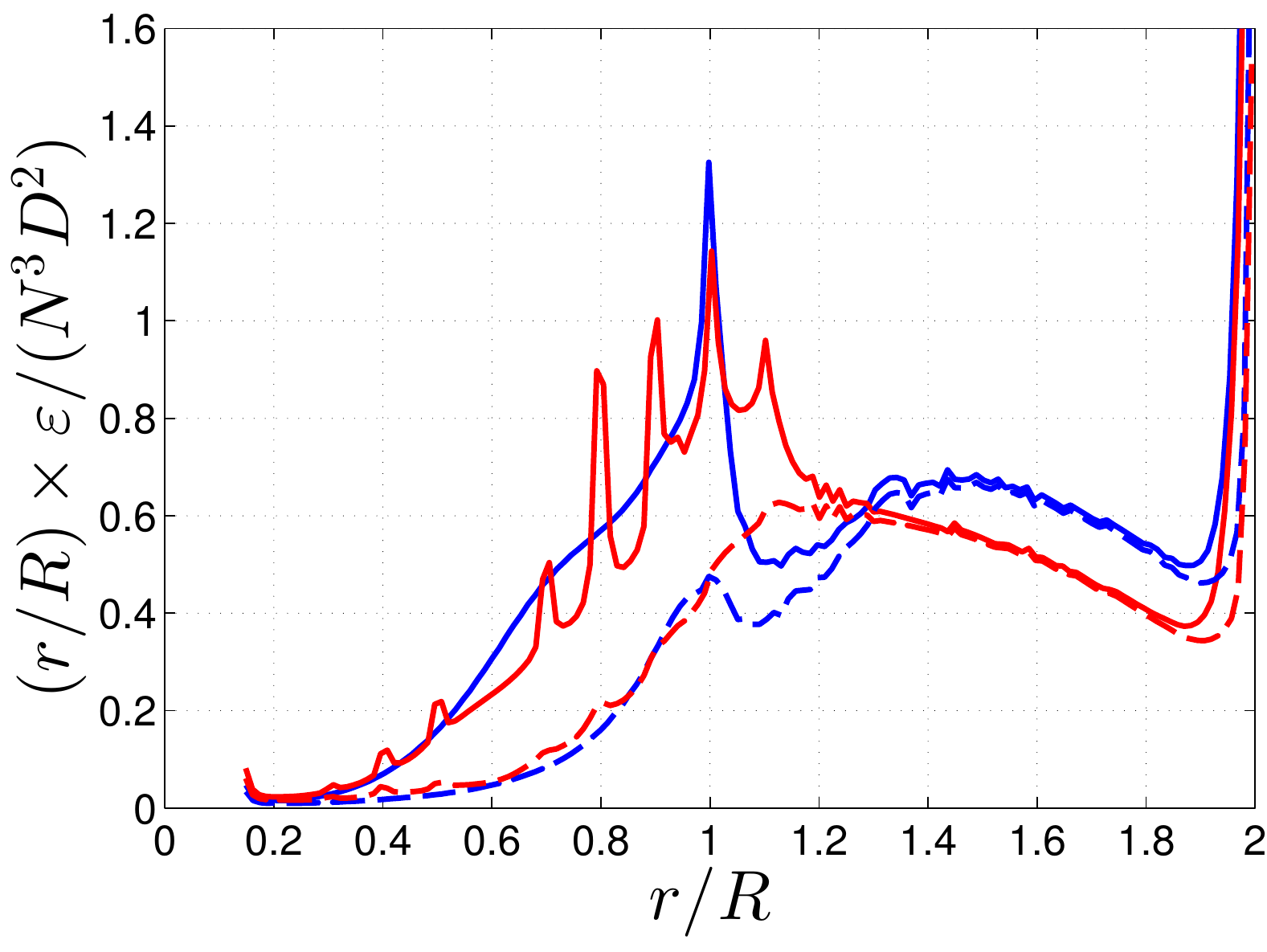}
        \caption{$-1/6<z/H<1/6$}
    \end{subfigure}  
  \caption{Radial profiles of dissipation obtained by averaging over
    concentric cylindrical surfaces with varying radii at $Re=1600$
    for both regular (reg.) and fractal (frac.) impellers, (a) for the
    entire height of tank, (b) only over the middle one third of the
    tank height. The averaged dissipation is normalized with $N^3 D^2$
    and weighted with $r/R$.}
  \label{fig:diss_profile}
\end{figure}

The dissipation of the total kinetic energy ($\varepsilon_K$) can be
decomposed in two parts: $\varepsilon_K=\varepsilon_T+\varepsilon_M$,
where $\varepsilon_T=2 \nu \left< s'_{ij} s'_{ij} \right>$ and
$\varepsilon_M=2 \nu S_{ij} S_{ij}$ are the turbulent and mean
velocity dissipation, respectively ($s'_{ij}=\frac{1}{2} \left(
\frac{\partial u'_i}{\partial x_j}+\frac{\partial u'_j}{\partial x_i}
\right)$ and $S_{ij}=\frac{1}{2} \left( \frac{\partial U_i}{\partial
  x_j}+\frac{\partial U_j}{\partial x_i}\right)$ are strain rate
tensors based on fluctuating velocity gradients and mean velocity
gradients, respectively). At $Re=320$, the share of $\varepsilon_T$ in
$\varepsilon_K$ integrated over the tank is only around 32\%, whereas
this value rises to 61\% at $Re=1600$. Figure \ref{fig:diss_profile}
illustrates radial profiles of $\varepsilon_K$ and $\varepsilon_T$
calculated by averaging over concentric cylindrical
surfaces with varying radii, for both impeller types at $Re=1600$. The
values are normalized with $N^3 D^2$. The calculation is performed for
the entire tank height (see Figure \ref{fig:diss_profile}a) and for
the middle one third of the tank height (see Figure
\ref{fig:diss_profile}b). The latter part of the tank ($-1/6<z/H<1/6$)
is significant, because this is where ca. 66\% of the total
dissipation and 75\% of the turbulent dissipation occurs (for
$Re=1600$). In Figure \ref{fig:diss_profile} the radial profiles are
weighted with $r/R$ to emphasize the increasing share of a cylindrical
surface in the total volume as its radius increases.

It can be seen that $\varepsilon_K$ is dominated by $\varepsilon_M$
between the shaft and the blade tip ($r/R \leqslant 1$). For the
regular impeller, it increases monotonically until a sharp peak at
$r/R=1$ while it presents multiple spikes for the fractal impeller,
one spike for every axial edge of the blade (refer to Figure
\ref{fig:cp}b for the shape of the fractal blade). The turbulent part
of the dissipation grows significantly in the radial range between
$0.8<r/R<1.3$ where trailing vortices emerge. In the bulk of the flow,
$1.3<r/R<1.9$, $\varepsilon_T$ accounts for almost the entire
dissipation. Especially in Figure \ref{fig:diss_profile}b representing
the middle one third of the tank height, the difference between
$\varepsilon_K$ and $\varepsilon_T$ is imperceptible. This is also the
part of the flow field where the blue curve (indicating the regular
impeller) is consistently higher than the red curve (denoting the
fractal impeller). The maximum values of the profiles of both
$\varepsilon_K$ and $\varepsilon_T$ are observed at the tank wall
$r/R=2$, that is ca. twice as high as the peak value seen at $r/R=1$
(not visible in Figure \ref{fig:diss_profile}). However, the high
level of dissipation near the wall is confined within a volume of very
small radial extent.

\begin{figure}[ht]
  \centering   
  \includegraphics[width=0.55\textwidth]{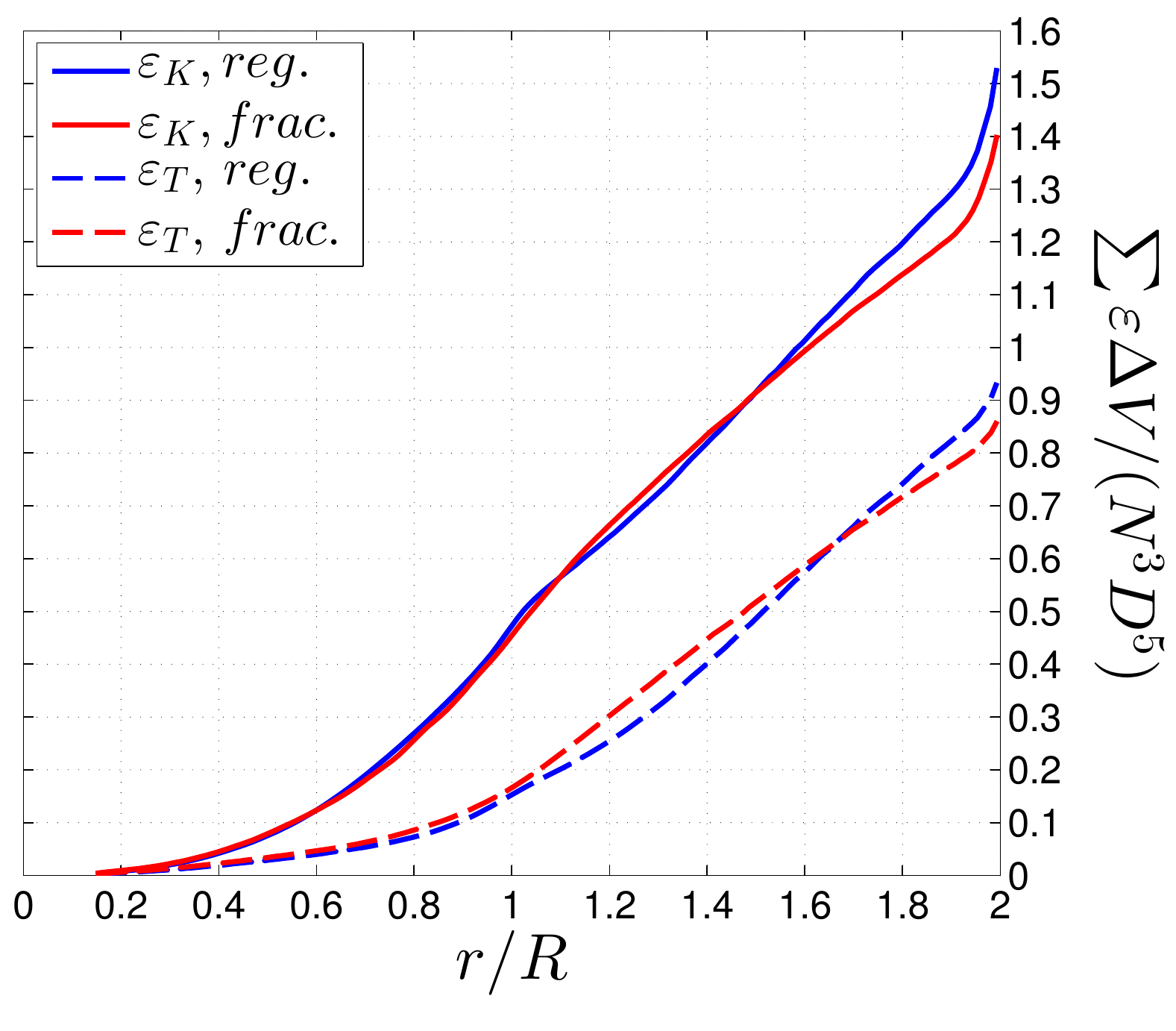} 
  \caption{Cumulative integrals of $\varepsilon_K$ and $\varepsilon_T$ at $Re=1600$ for both impeller types.}
  \label{fig:diss_integral}
\end{figure}

Figure \ref{fig:diss_profile} suggests that the 8\% difference in the
total dissipation between the two impeller types is mainly due to the
turbulent dissipation observed in the radial range $1.3<r/R<1.9$. In
order to further substantiate this conclusion, the profiles of
$\varepsilon_K$ and $\varepsilon_T$ are volume-integrated. Figure
\ref{fig:diss_integral} shows at every radial position ($r/R$) the
volume-integrated dissipation from 0 to this point for the entire
height of the tank. The resulting value is normalized with $N^3 D^5$
so that it corresponds to the power numbers listed in Table
\ref{table:all_power}, when the curves representing the integral of
$\varepsilon_K$ reach $r/R=2$.

Figure \ref{fig:diss_integral} shows that the cumulative integrals of
$\varepsilon_K$ are equal for both impeller types up to ca. $r/R=1.1$,
i.e. the tip of the fractal blade. From this point until ca. $r/R=1.9$
the straight and dashed lines continue almost parallel to each other,
indicating that any increase in the integral of $\varepsilon_K$ is
mainly due to $\varepsilon_T$. The integral of turbulent dissipation
until ca. $r/R=1.3$ is higher for the fractal impeller case, which
implies that the smaller trailing vortices in the wake of the fractal
blade dissipate over a shorter distance from the impeller. From
$r/R=1.3$ to ca. $r/R=1.9$, the blue dashed line has a higher slope
than the red dashed line and this is where the surplus of dissipation
in the regular impeller case is observed with respect to the fractal
impeller case. Finally, the curves present a sharp increase in the
near-wall region, but the difference between the blue and red curves
does not change over this last part. It can be concluded that the two
large coherent structures in the wake of the regular blade dissipate
further from the impeller, in other words survive longer before
decaying, and remove a higher amount of energy during this process.

\begin{figure}[ht] 
  \centering      
    \begin{subfigure}[]{0.52\textwidth}
    \centering   
    \includegraphics[width=\textwidth]{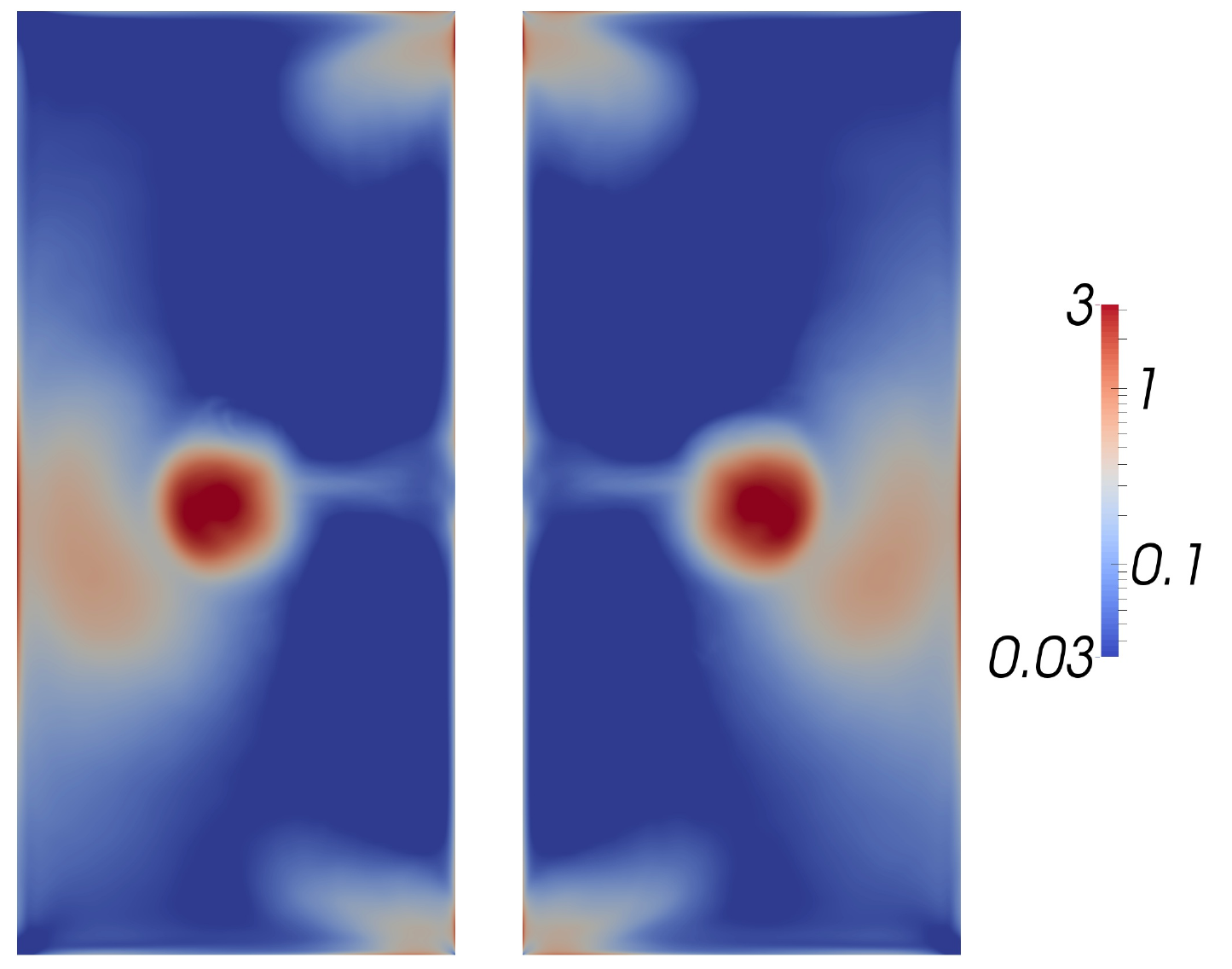}   
    \caption{}
    \end{subfigure}   
    \begin{subfigure}[]{0.41\textwidth}
    \centering 
    \includegraphics[width=\textwidth]{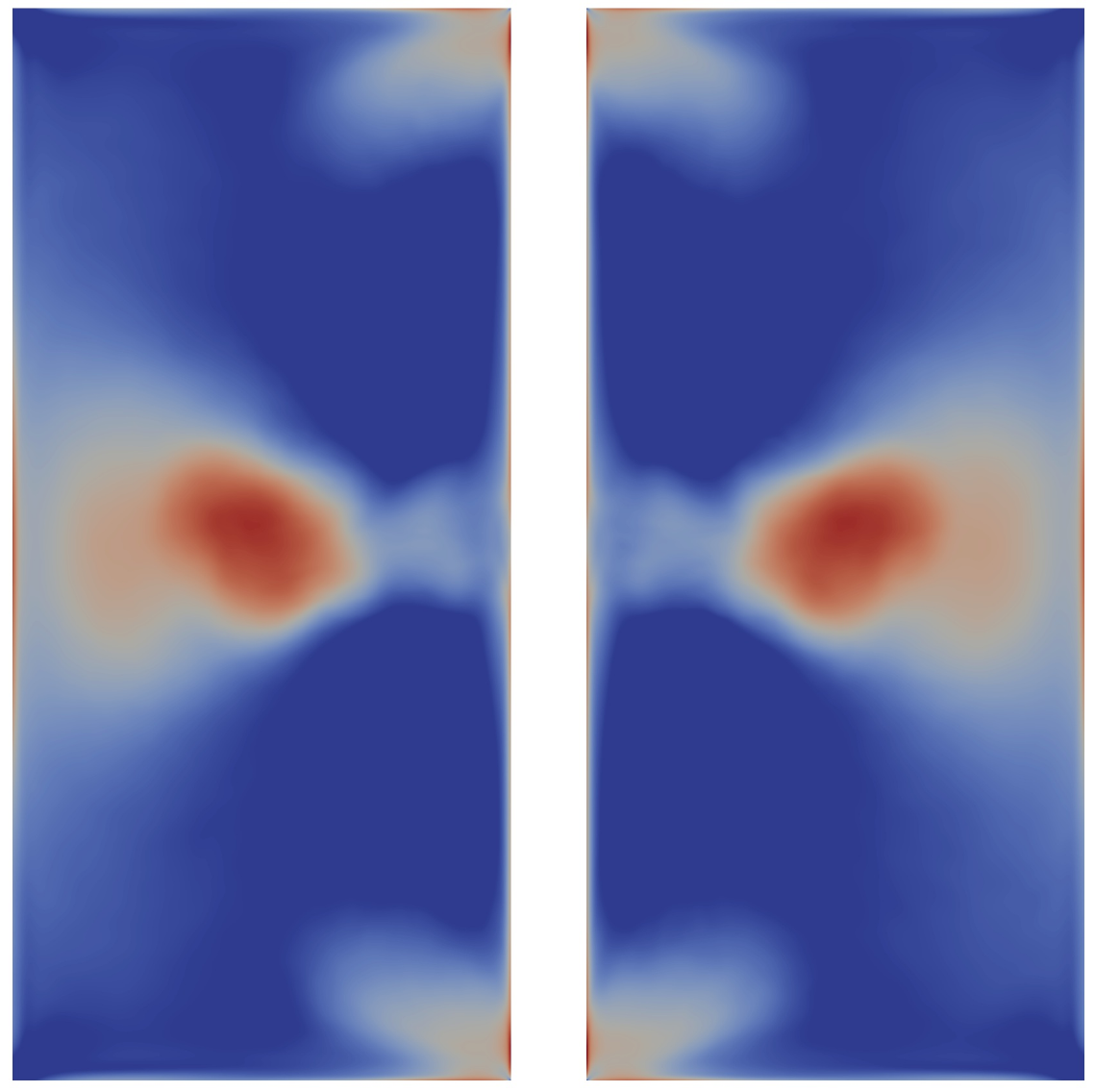}            
    \caption{}
    \end{subfigure}   
  \caption{Contours of $\varepsilon_T/(N^3 D^2)$ at $Re=1600$ at 30\textdegree\ downstream of the blade wakes for (a) regular, (b) fractal impeller.}
  \label{fig:diss_contour}
\end{figure}

The above argument is made more clear when we examine contour plots of
the turbulent dissipation $\varepsilon_T$ in the vicinity of the
trailing vortices. These are displayed in Figure
\ref{fig:diss_contour} over a vertical cross-section of the entire
domain at 30\textdegree\ downstream of the blades for $Re=1600$. The
highest values are observed in the region where trailing vortices emerge. 
The higher level of $\varepsilon_T$ in the wake of the regular impeller is 
notable with the darker red colour when compared to the fractal impeller case; 
the latter has a rather dispersed distribution of $\varepsilon_T$. The
intermediate values of $\varepsilon_T$ between the wall and the vortex
cores mark the wake of the succeeding blades. This quantity drops
quickly at least two orders of magnitude as the fluid moves in the
(positive or negative) axial direction further from the mid-height of
the tank.

\begin{figure}[ht]
  \centering   
  \includegraphics[width=0.6\textwidth]{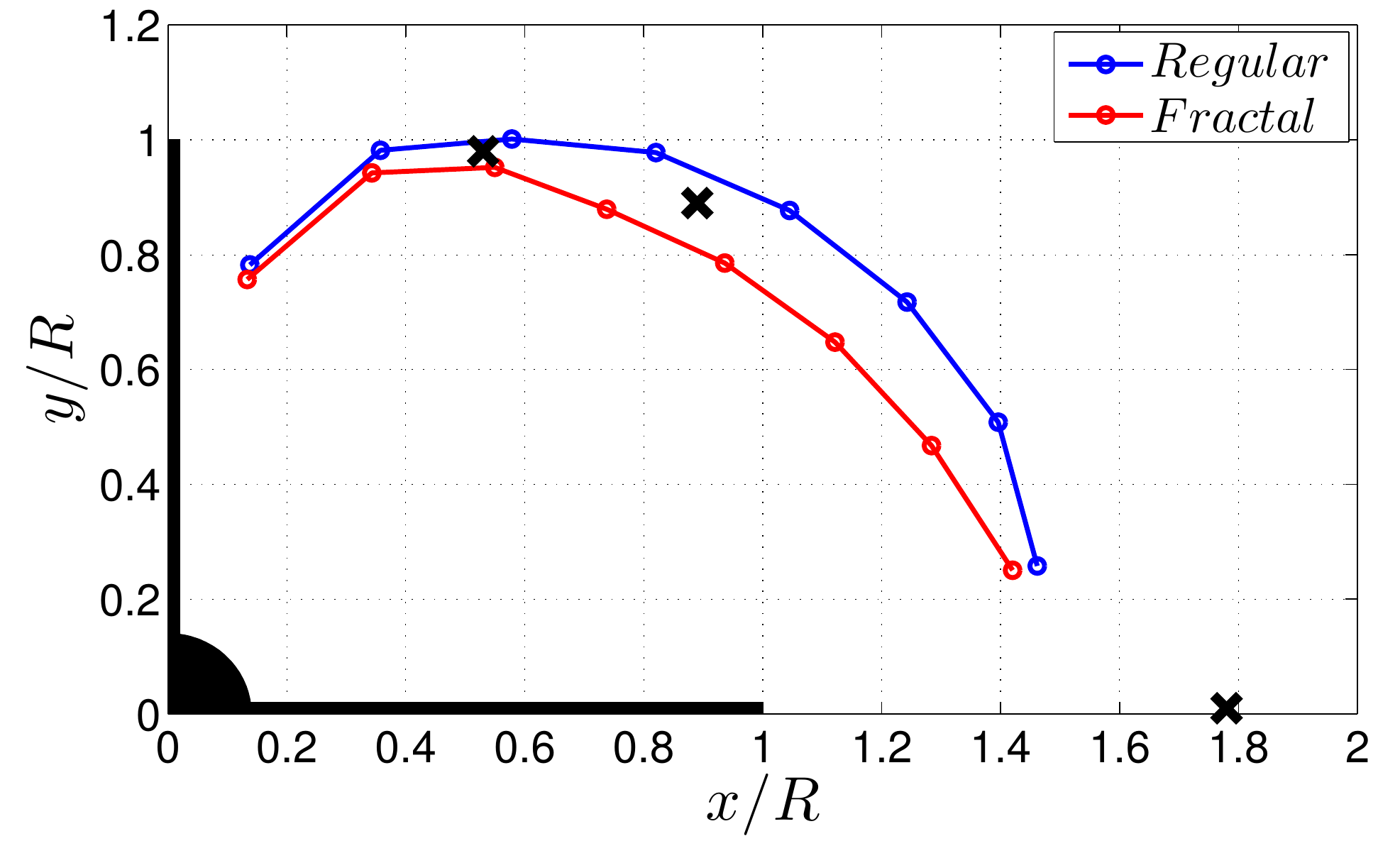} 
  \caption{Trailing vortex paths for both impeller types at $Re=1600$ are shown with the lines and the locations of probe points are marked with cross symbols. A sketch of the impeller blades improves the clarity of the figure.}
  \label{fig:vort_path}
\end{figure}

\begin{figure}[ht] 
  \centering   
    \begin{subfigure}[b]{0.32\textwidth}
    \centering   
        \includegraphics[width=\textwidth]{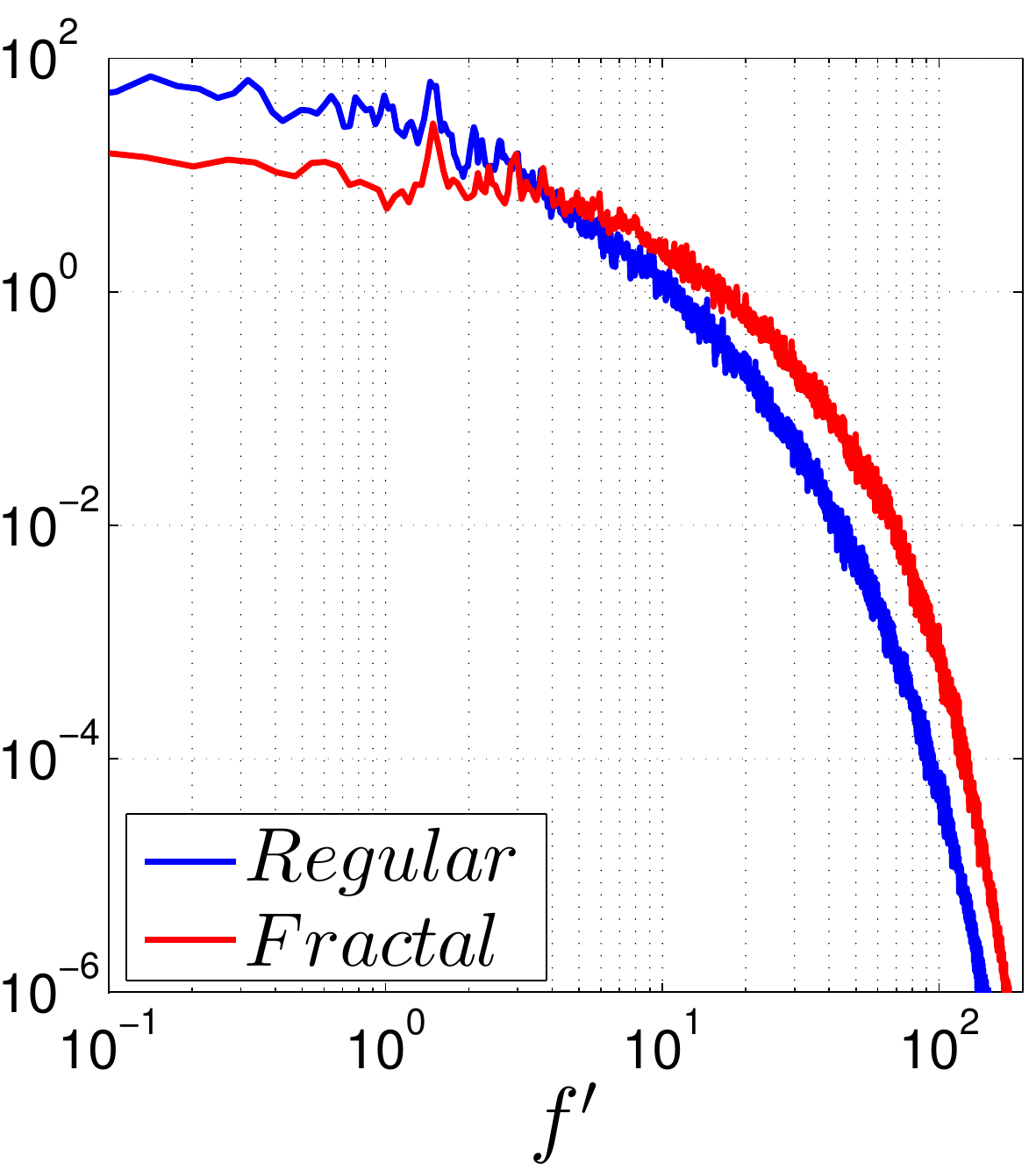}
        \caption{Probe-1}
    \end{subfigure}  
    \begin{subfigure}[b]{0.32\textwidth}
    \centering 
        \includegraphics[width=\textwidth]{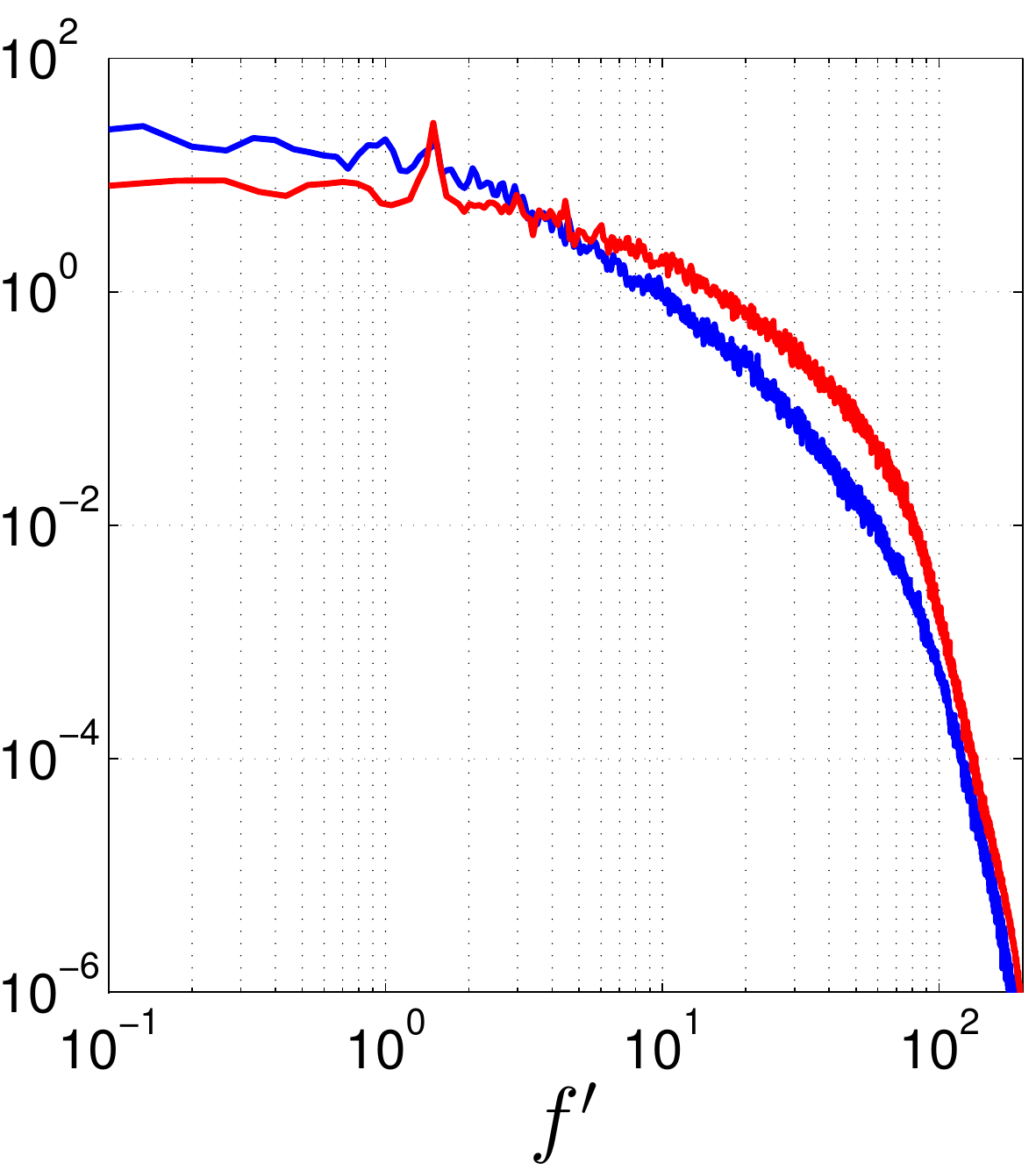}
        \caption{Probe-2}
    \end{subfigure}  
    \begin{subfigure}[b]{0.32\textwidth}
    \centering 
        \includegraphics[width=\textwidth]{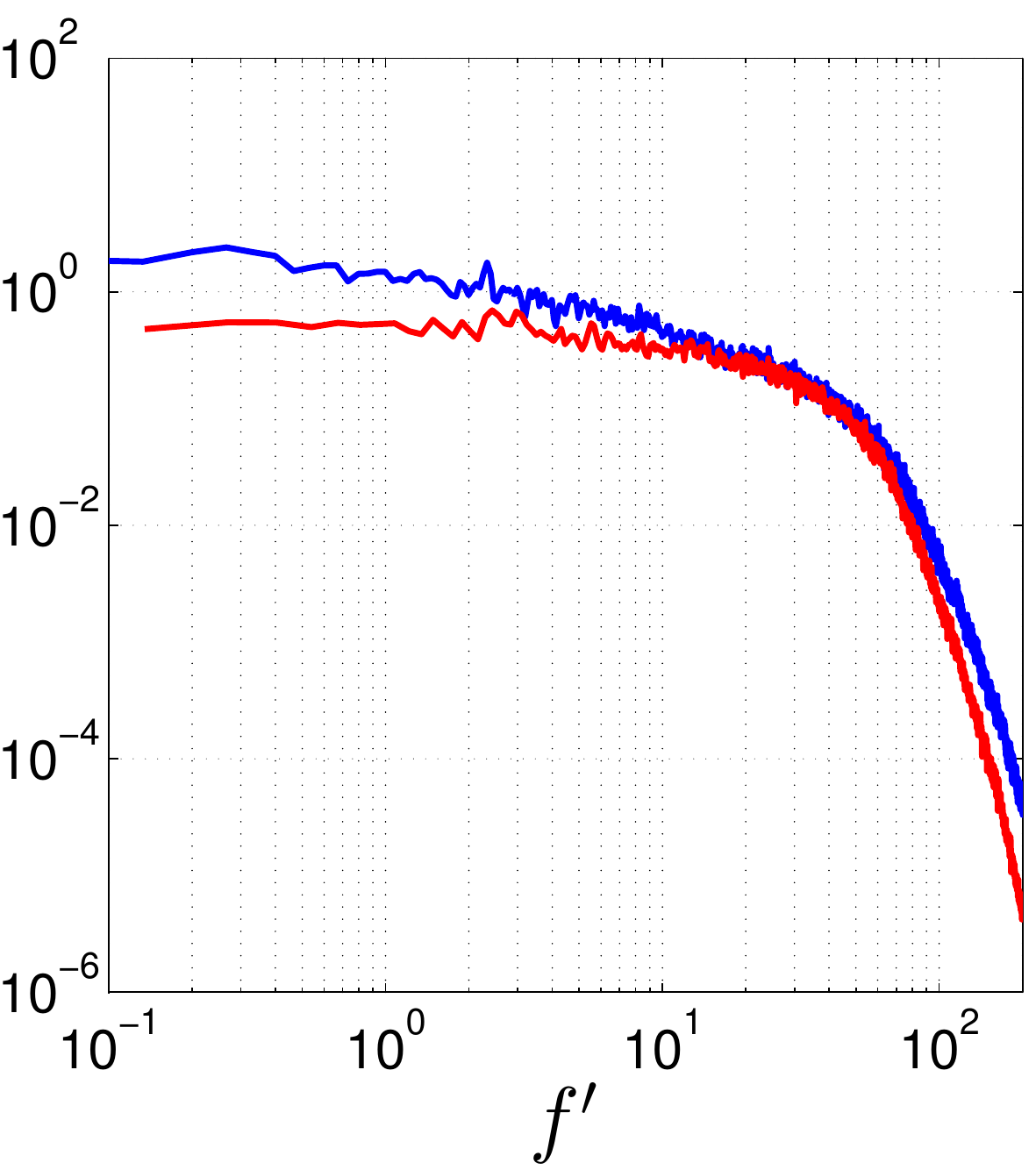}
        \caption{Probe-3}
    \end{subfigure}  
  \caption{PSDs of $\varepsilon_T(t)$ at the three probe points marked in Figure \ref{fig:vort_path} with cross symbols, for $Re=1600$.}
  \label{fig:diss_PSD}
\end{figure}

When $\varepsilon_T(t)=2 \nu s'_{ij} s'_{ij} $ is analysed in the frequency domain, further differences are observed in the wakes of regular and fractal blades. By dropping the averaging operation, the $\varepsilon_T(t)$ can be considered as a function of time. In order to evaluate this quantity, all nine velocity gradients $(\frac{\partial u_i}{\partial x_j})$ are recorded at three probe points on-the-fly and used to compute the components of the $s'_{ij}$ tensor. The locations of the aforementioned probes are illustrated in Figure \ref{fig:vort_path} with cross symbols, along with the paths of the vortex cores at $Re=1600$ for both impeller types. In order to show the vortex path, vortex cores are determined based on the $\lambda_2$-criterion \cite{Jeong1995} and the radial coordinate is averaged over their cross-section at various angles behind the blade. Also the schematic image of impeller blades is illustrated in the same figure. The probe points will be referred to as Probe-1,2,3; the numbering follows the increasing x-location displayed in Figure \ref{fig:vort_path}, i.e. with the increasing downstream distance away from the blade. The locations of Probe-1,2,3 in cylindrical coordinates are ca. $r/R=1.1,\,1.3,\,1.8$ at 30\textdegree, 45\textdegree\ and 90\textdegree\ behind the blade, respectively. In order to obtain the power spectral density (PSD) of $\varepsilon_T(t)$, the nine components of the $s'_{ij}$ tensor are normalized with $N$, their PSDs are computed separately and added in the frequency domain. These PSDs are illustrated in Figure \ref{fig:diss_PSD} over a nondimensional frequency $f'=f/N$, for both impeller types and evaluated at the aforementioned probe points.

When any one of the spectra in Figure \ref{fig:diss_PSD} is integrated over the frequency, the result is proportional to the value of $\left< s'_{ij} s'_{ij} \right>$ at this point. For instance, at Probe-2 shown in Figure \ref{fig:diss_PSD}b, the values of $\varepsilon_T$ of both cases are equal, hence the integrals of the blue and red curves. On the other hand, it is notable that the red curve is higher than the blue curve over frequencies in the range of ca. $3<f'<100$, whereas it is lower over the range of lower frequencies. This means that the flow field in the wake of a fractal blade has at this point (Probe-2) a surplus of dissipation for the fluctuations at higher frequencies with respect to the wake of a regular blade, which is balanced in the latter case with a surplus of dissipation for the fluctuations at lower frequencies. Moreover, at Probe-3 the turbulent dissipation is ca. 40\% higher for the regular impeller, mainly due to the difference at low frequencies as seen in Figure \ref{fig:diss_PSD}c. This point is in the part of the flow, where the radial profile of the dissipation illustrates a significant difference between the regular and fractal impeller cases, as shown earlier in Figure \ref{fig:diss_profile}. In addition to the region in the flow field, which accounts for the surplus of dissipation in the regular impeller case, it is also demonstrated now that this surplus is mainly due to the difference of the fluctuations at low frequencies. 

\begin{figure}[ht] 
  \centering   
    \begin{subfigure}[b]{0.49\textwidth}
    \centering   
        \includegraphics[width=0.65\textwidth]{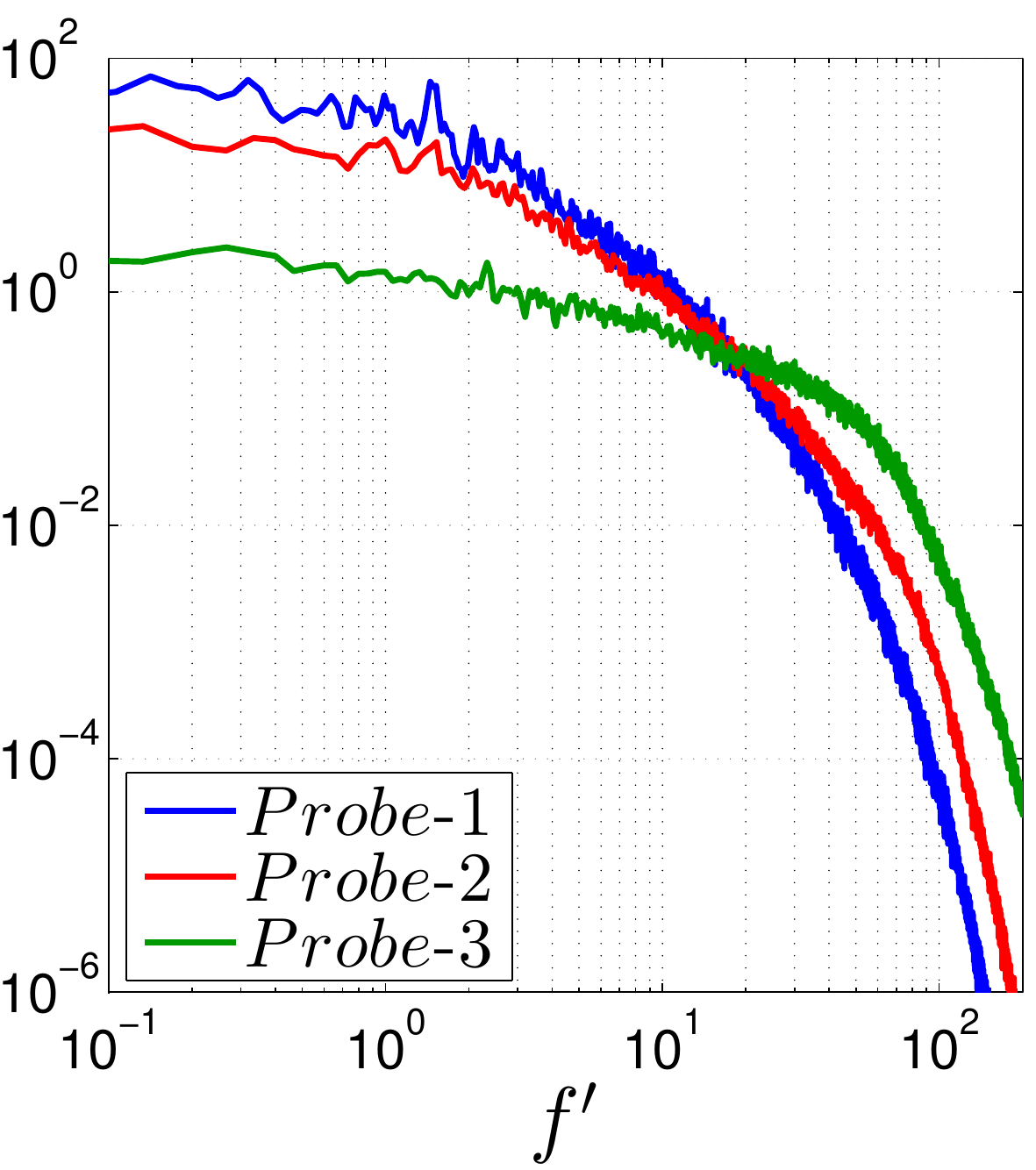}
        \caption{Regular impeller.}
    \end{subfigure}  
    \begin{subfigure}[b]{0.49\textwidth}
    \centering 
        \includegraphics[width=0.65\textwidth]{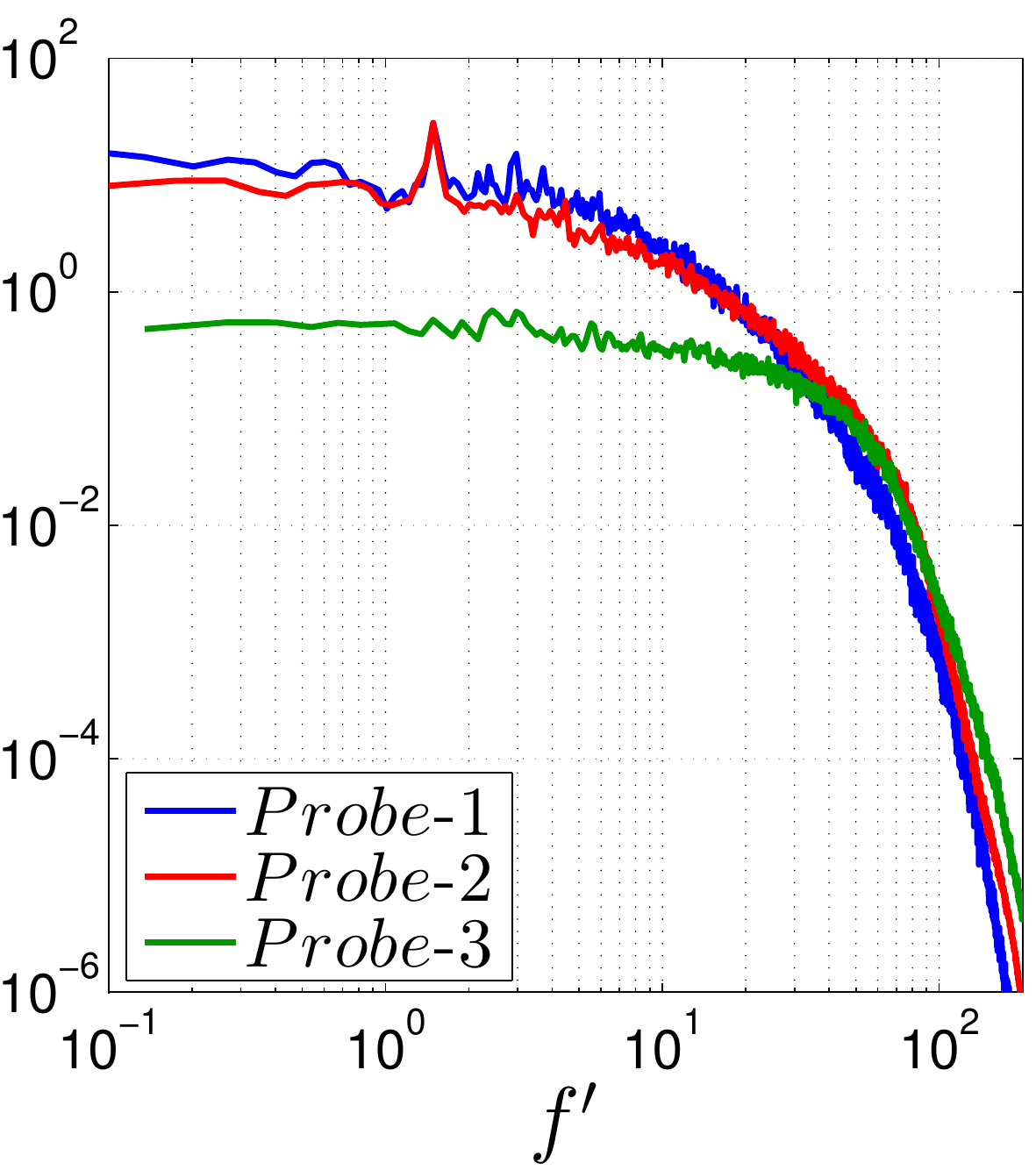}
        \caption{Fractal impeller.}
    \end{subfigure}  
  \caption{PSDs of $\varepsilon_T(t)$ at the three probe points compared separately for regular and fractal impeller.}
  \label{fig:diss_PSD2}
\end{figure}

In order to illustrate the change in spectra along the vortex paths, Figure \ref{fig:diss_PSD2} compares the PSDs of $\varepsilon_T(t)$ at the three probe points, separately for both impeller types. As the fluid moves further from the blade, there is a substantial decay in the spectral density at low frequencies (for ca. $f'<20$) in both cases. Meanwhile for the regular impeller, the spectral density increases for the part at high frequencies, i.e. $f'>20$. This can be a shift of the dissipation from lower frequencies partially to higher frequencies. If we assume that the dissipation at higher frequencies is mainly due to vortical structures at smaller scales, we may interpret that this shift is due to the cascade of the energy from large to small scales. It might be the case that the energy reaches down to smallest scales only after a certain time and distance from the regular blade, due to the large initial size of vortices. Hence, at Probe-3 there is an increased dissipation for $f'>20$, compared to Probe-1 and 2, see Figure \ref{fig:diss_PSD2}a. On the other hand, for the fractal impeller this part of spectra $(f'>20)$ does not change much along the vortex path, see Figure \ref{fig:diss_PSD2}b, and is very similar to what is seen at Probe-3 of the regular impeller. This part might have developed much earlier in terms of time and distance for the fractal impeller, due to the smaller length scale of vortices. Nevertheless, this interpretation requires the support of spectra in the scale-space before drawing firm conclusions. 

Using the assumptions of the last paragraph, the difference at low frequencies seen in Figure \ref{fig:diss_PSD}c, which leads to 40\% higher dissipation for the regular impeller at Probe-3, can be linked to the larger size of its trailing vortices. This would be in accordance with the earlier observations, that the larger trailing vortices decay further from the impeller and dissipate higher energy. 

In conclusion, the same mechanism leading to a higher drag coefficient of the blades and the higher transport of angular momentum, is also responsible for the dissipation of the higher power drawn by the regular impeller. The modification in the blade shape is able to significantly change the trailing vortex structures, hence alter the impeller power as well the dissipation in the flow field. This result is a promising example of how such intrinsic properties might be tuned by means of modifications on the blade shape, for instance to match specific process requirements.

\section*{Summary and Conclusions}

Previous experiments\cite{Steiros} have shown that the fractal impeller has $11-12\%$ reduced torque compared to the regular impeller at $Re=1-2\times10^5$. A similar difference of $8\%$ was also found in the DNS results\cite{Basbug} for a lower Reynolds number ($Re=1600$).  In order to explain the origin of this difference, we conducted a detailed analysis of the flow inside an unbaffled tank stirred with regular and fractal impellers using the DNS data.

Firstly, it is observed that the drag coefficient of the fractal blade is distinctly lower than that of the regular blade at $Re=1600$ (in agreement with experiments at higher $Re$\cite{Steiros}), leading to the torque reduction. In order to analyse this outcome, the relation between the pressure distribution on the blades and the velocity field around the blades is investigated. It is demonstrated that profiles of $V_\theta$ upstream of both blade types are close to each other; the main difference is in the wake of the blades. The volume of the flow separation region is $7\%$ smaller, and the maximum magnitude of $V_\theta$ towards the suction side is ca. $50\%$ lower in the wake of the fractal blade compared to the wake of the regular blade. These differences emerge since the concave edges of the fractal blade allow the upcoming fluid to penetrate into the separation region. It is also shown that the recirculation pattern on the suction side is directly connected with the generation of trailing vortices. While there are two coherent structures behind a regular blade, the fractal blade has multiple and smaller trailing vortices.

Since any difference in the impeller torque is directly reflected in the transport of angular momentum from the impeller to the fluid, this quantity was also compared between the two blade types at $Re=1600$, using a control volume (CV) around the impeller. First, it is noted that the fractal impeller yields a $10\%$ higher mass flow rate through the borders of the CV, despite having $8\%$ lower $N_p$. The net transport of angular momentum per unit flow rate is $18\%$ lower for the fractal impeller. This aspect outweighs its $10\%$ higher flow rate and explains the $8\%$ lower transport of angular momentum with respect to the regular impeller. Elaborating further, the radial advective transport per unit flow rate out of the CV is decomposed into mean and turbulent transport, i.e. $l_{out,mean}$ based on $r U_{\theta} U_r$ and $l_{out,turb}$ based on $r\left< u'_{\theta} u'_r \right>$, respectively. Although $l_{out,mean}$ has a larger contribution in the total $l_{out}$, the difference between the regular and fractal impellers is mainly due to the role of $l_{out,turb}$ which is ca. $30\%$ higher for the regular impeller. Furthermore, the three dimensional isosurfaces of $r\left< u'_{\theta} u'_r \right>$ coincide exactly with trailing vortices, showing that these structures are responsible for the turbulent transport. It can be inferred that the alteration of trailing vortices is the link between the reduction in the torque and the reduction in the transport of angular momentum.

As the power draw is $8\%$ lower for the fractal impeller, so must be the integral dissipation, as well. This quantity includes the contribution of mean velocity gradients ($\varepsilon_M$) and the turbulent dissipation ($\varepsilon_T$). It is noted that $\varepsilon_T$ is concentrated in the vicinity of trailing vortices and it makes up $61\%$ of the total dissipation of the entire tank, at $Re=1600$. It is observed that the reduction of $8\%$ in the total dissipation, when the fractal impeller is compared to the regular impeller, is mainly due to the difference in $\varepsilon_T$ in the radial range of $1.3<r/R<1.9$. The frequency distribution of the dissipation was assessed at 3 probe points along the vortex paths. It is  illustrated that the wake of the regular impeller contains higher dissipation compared to the wake of the fractal impeller due to fluctuations at low frequencies $(f'<20)$, which accounts for the difference in the bulk of the flow ($1.3<r/R<1.9$). 

It is concluded that by means of modifications in the blade shape it is possible to influence crucially the trailing vortex structures. Consequently, the drag coefficient of the blades, the impeller torque and the energy dissipation characteristics can be altered, in order to tune these according to process requirements and optimise the impeller design. In the case of fractal impeller, the reduction of $N_p$, the increase in $N_q$ and the equal level of integrated turbulent kinetic energy with respect to the regular impeller are promising findings for an improved process efficiency.

\section*{Appendix: Time-lag of the energy dissipation}

\renewcommand{\thefigure}{A\arabic{figure}}
\setcounter{figure}{0}

\begin{figure}[ht] 
  \centering   
\includegraphics[width=0.6\textwidth]{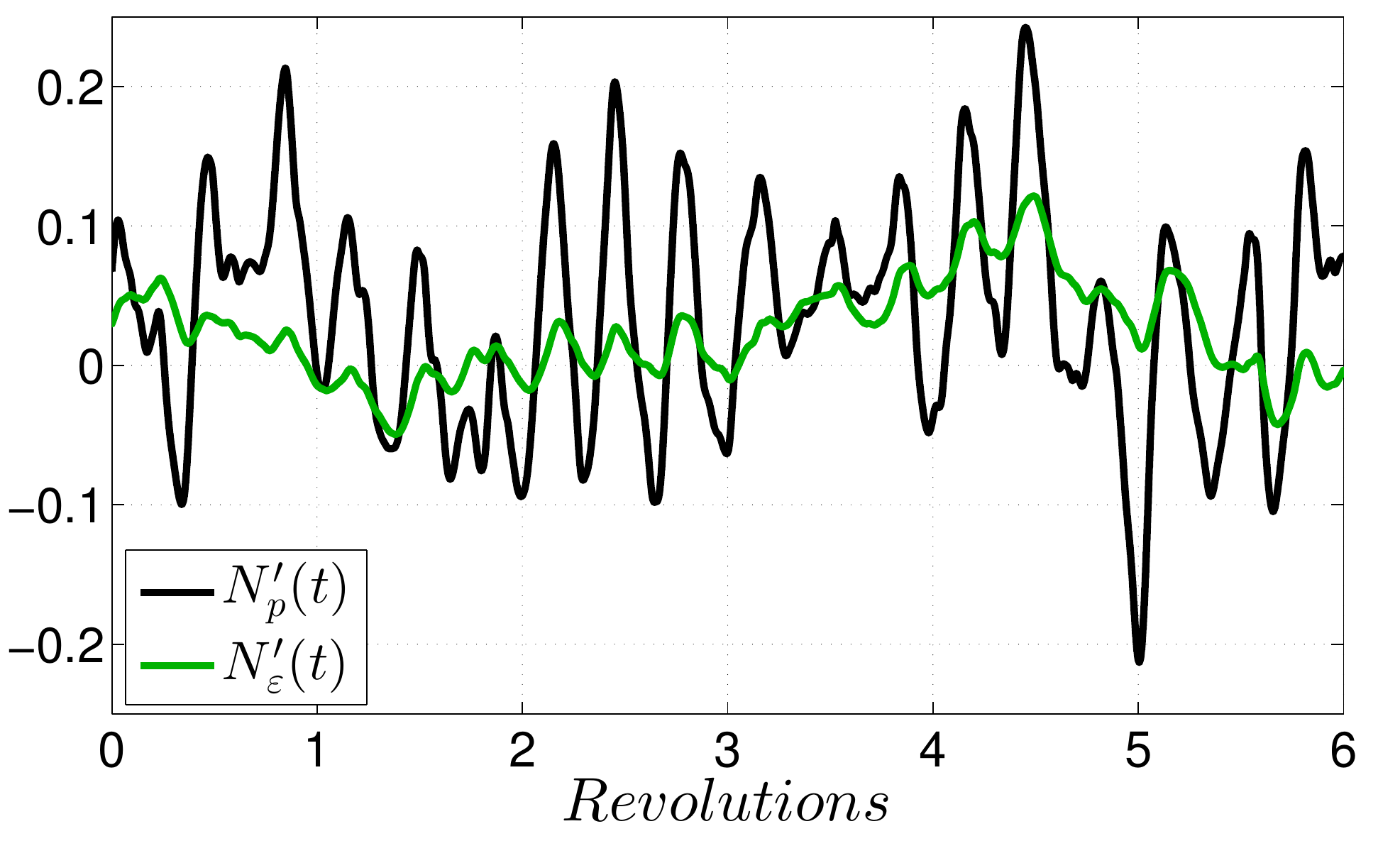}
  \caption{Fluctuations of power consumption are plotted for fractal
    impeller at $Re=1600$. $N_p'(t)$ is based on the power drawn by
    the impeller and $N_{\varepsilon}'(t)$ is based on the volume
    integral of the total dissipation over the tank volume.}
\label{fig:np_fluct}
\end{figure}

The time-average values of the impeller power and integral dissipation
have been shown to be in balance, except for a numerical dissipation
of ca. $4\%$ at $Re=1600$ (see Table \ref{table:all_power}). The
fluctuations in time of the power number, obtained from pressure
integration on the impeller surface (denoted by $N_p'(t)$), 
and of the volume integral of the instantaneous dissipation 
(denoted by $N_{\varepsilon}'(t)$) are plotted in Figure \ref{fig:np_fluct} 
against the number of revolutions.
          
\begin{figure}[ht] 
  \centering        
    \begin{subfigure}[]{0.49\textwidth}
    \centering 
    \includegraphics[width=\textwidth]{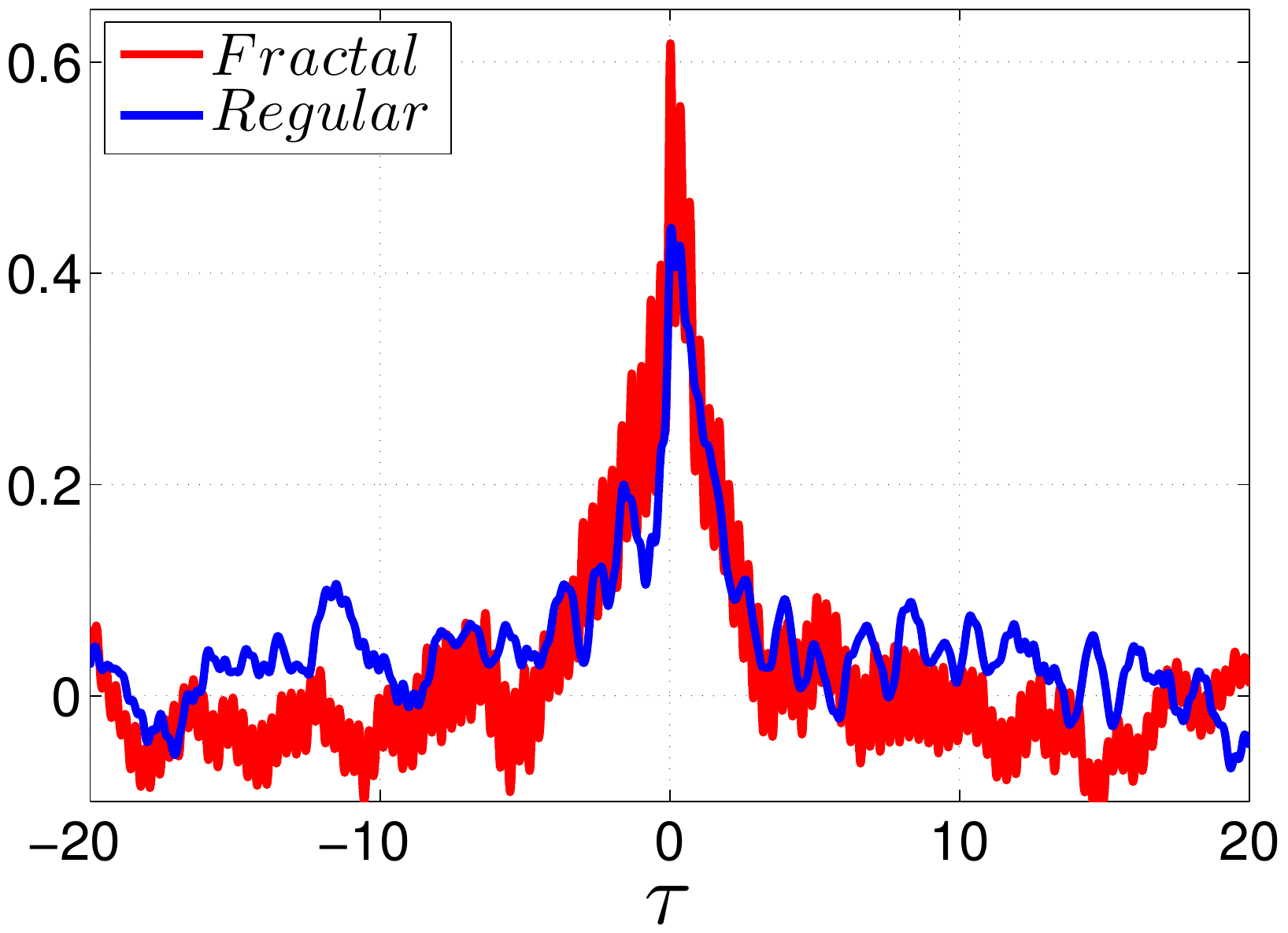}            
    \caption{}
    \end{subfigure}     
    \begin{subfigure}[]{0.49\textwidth}
    \centering 
    \includegraphics[width=\textwidth]{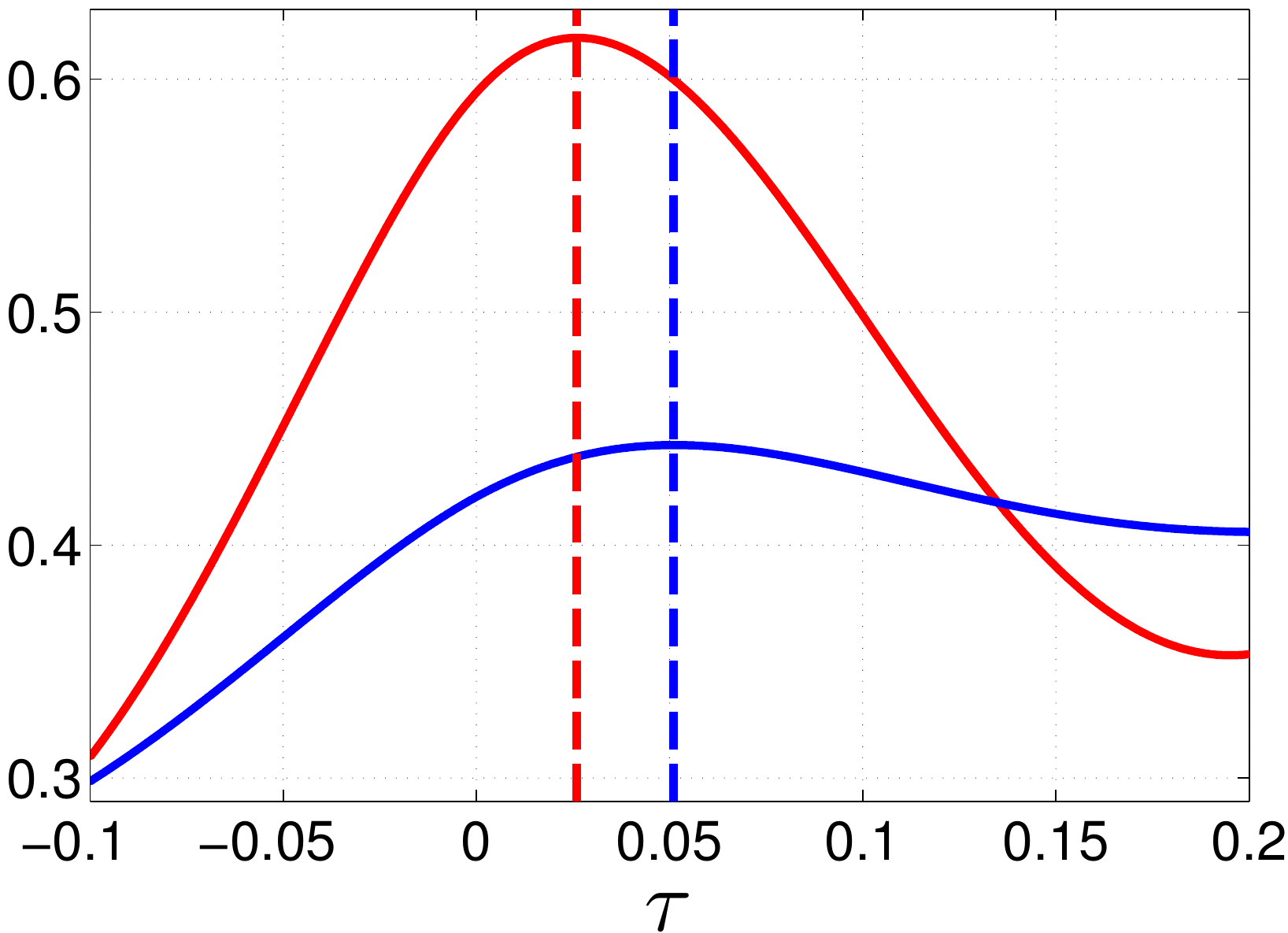}            
    \caption{}
    \end{subfigure}     
  \caption{Time-correlation $Cor(\tau)$, defined in Equation
    \ref{eq:corr}, plotted against $\tau$. a) Comparison between
    regular and fractal impellers, b) zoom on the peaks.}
  \label{fig:np_cross}
\end{figure}

It is seen in Figure \ref{fig:np_fluct} that $N_p'(t)$ presents strong
fluctuations with multiple peaks per revolution. As discussed in our
previous paper,\cite{Basbug} these fluctuations are directly linked to
the unsteady motion of trailing vortices and up-and-down swinging
motion of the radial jet, at a frequency $f'=3$ at $Re=1600$, for both
regular and fractal impellers. The amplitude of these fluctuations are
weaker for $N_{\varepsilon}'(t)$ (green curve). The instantaneous
difference between impeller power and integral dissipation is equal to
the time derivative of the kinetic energy integrated over the tank
volume, as the energy balance dictates. A detail, admittedly difficult
to notice in Figure \ref{fig:np_fluct}, is that there is a time-lag
between the peaks of the two quantities. In other words, the peaks in
the dissipation follow the peaks in the impeller power with a certain
time delay. This delay can be evaluated with the help of the two-time
cross-correlation function between the two signals:
\begin{equation}
Cor(\tau)= \frac{\left< N_p'(t) \, N_{\varepsilon}'(t+\tau) \right>}{ \sqrt{ \left< N_p'(t)^2 \right> \, \left<N_{\varepsilon}'(t)^2 \right>} } \:\:,  
\label{eq:corr}
\end{equation}
where angular brackets $\left< \right>$ represent the time-averaging
operation. The correlation $Cor(\tau)$ is plotted in Figure
\ref{fig:np_cross}a for both impeller types. The value of $\tau$, at
which $Cor(\tau)$ attains the global maximum value characterizes the
time-lag between $N_p'(t)$ and $N_{\varepsilon}'(t)$. A closer look at
the peaks is provided in Figure \ref{fig:np_cross}b. These peaks
appear at $\tau_r=0.051$ for the regular impeller and $\tau_f=0.026$
for the fractal impeller. In Figure \ref{fig:np_cross}b, these are
indicated with vertical dashed lines in the same color as the
corresponding curves. It is noteworthy that $\tau_r$ is twice as large
as $\tau_f$.

These time-lags characterise the time required for the kinetic energy
to cascade from the injection length-scales to the smallest scales
where it dissipates. In our cases, highest turbulence production is
observed in trailing vortex cores and these represent the most
energetic scales in the flow. Taking this into account, the time-lag
of the dissipation may be longer for the regular impeller due to the
larger scales at which the energy is injected. It is also remarkable,
though, that $Cor(\tau)$ presents a much more distinct peak for the
fractal impeller than for the regular one. The fractal impeller seems
to sharpen the cascade time around a particular value whereas the
range of cascade times appears to be quite wide for the regular
impeller. This observation will require a dedicated future study of
its own.

\section*{Acknowledgements}

The authors acknowledge the EU support through the FP7 Marie Curie
MULTISOLVE project grant no. 317269, the computational resources
allocated in ARCHER HPC through the UKTC funded by the EPSRC grant
no. EP/L000261/1 as well as the CX2 facility of Imperial College
London. JCV also acknowledges ERC Advanced Grant 320560. The authors
are grateful to Konstantinos Steiros for providing the experimental
data and for helpful discussions. SB also acknowledges Felipe Alves Portela 
and Nikitas Thomareis for their help in the solution of computational issues 
in the initial phase of the project.

\renewcommand\refname{Literature cited}

\bibliography{library}
\bibliographystyle{aichej}

\end{document}